\documentclass{aa} 
\usepackage{graphicx}
\usepackage{natbib}

\newcommand{\beq}{\begin{equation}}
\newcommand{\bea}{\begin{eqnarray}}
\newcommand{\eeq}{\end{equation}}
\newcommand{\eea}{\end{eqnarray}}
\newcommand{\p}{\partial}
\newcommand{\BB}{{\bf B}}
\newcommand{\UU}{{\bf u}}
\newcommand{\XX}{{\bf X}}

\newcommand{\pol}{polo\"{\i}dal}
\newcommand{\D}{\Delta}
\begin{document}
\title{Relativistic particle transport in extragalactic jets}
\subtitle{I. Coupling MHD and kinetic theory}
\titlerunning{Relativistic particle transport in extragalactic jets}
\authorrunning{Casse \& Marcowith}
\author{Fabien Casse\inst{1} and Alexandre Marcowith\inst{2}}
\offprints{A. Marcowith}
\institute{FOM-Institute for Plasma physics ``Rijnhuizen'', PO Box 1207
NL-3430 BE Nieuwegein, The Netherlands \email{fcasse@rijnh.nl}
\and C.E.S.R., 9 avenue du colonel Roche, BP 4346, F-31028 Toulouse, France \email{Alexandre.Marcowith@cesr.fr}}

\date{Received December 24th, 2002; Accepted March 26th, 2003}

\abstract{Multidimensional magneto-hydrodynamical (MHD) simulations coupled
with stochastic differential  equations (SDEs) adapted to test particle
acceleration and transport in complex astrophysical flows are
presented. The numerical scheme allows the investigation of shock
acceleration, adiabatic and radiative losses as well as diffusive spatial
transport in various diffusion regimes. The applicability  of SDEs to
astrophysics is first discussed in regards to the different regimes and the
MHD code spatial resolution.   The procedure is then applied to 2.5D
MHD-SDE simulations of kilo-parsec scale extragalactic jets. The ability of
SDE to reproduce analytical solutions of the diffusion-convection equation
for electrons is tested through the incorporation of an increasing number
of effects: shock acceleration, spatially dependent diffusion coefficients
and synchrotron losses. The SDEs prove to be efficient in various shock
configuration occurring in the inner jet during the development of the
Kelvin-Helmholtz instability. The particle acceleration in snapshots of
strong single and multiple shock acceleration including realistic spatial
transport is treated. In chaotic magnetic diffusion regime, turbulence
levels $\eta_T=<\delta B^2>/(B^2+<\delta B^2>)$ around $0.2-0.3$  are found
to be the most efficient to enable particles to reach the highest
energies. The spectrum, extending from 100 MeV to few TeV (or even 100 TeV
for fast flows), does not exhibit a power-law shape due  to transverse
momentum dependent escapes. Out of this range, the confinement is not  so
efficient and the spectrum cut-off above few hundreds of GeV, questioning
the Chandra observations of X-ray knots as being synchrotron radiation.
The extension to full time dependent simulations to X-ray extragalactic
jets is discussed.  
}

\maketitle
\section{Introduction}
Extragalactic jets in radio-loud active galactic nuclei (AGN) show
distinct, scale dependent structures. At parsec (pc) scales from the core,
superluminal motions have been detected  using VLBI technics. The jets
decelerate while reaching kiloparsec (kpc) scales and power large scale
luminous radio lobes. The inner physical conditions are still widely
debated. Main uncertainties concern bulk velocities, matter content,
emission and acceleration mechanisms, the way energy is shared between
magnetic field and plasma and finally effects of the turbulent flow on
relativistic particles. \\  
Recent X-ray high resolution observations by
Chandra, combined with Hubble space telescope (HST) and radio data allow
unprecedented multi-wavelength mapping of the jet structures which lead to
improved constraints of the physics \citep{Setal02}. The kpc jets show
nonthermal radio and optical spectra usually associated with synchrotron
radiation produced by highly relativistic TeV electrons  (positrons may
also contribute to the flux). The origin of the X-ray emission is more
controversial and could result from synchrotron radiation or Inverse
Compton (IC)  re-processing of low energy photons coming from different
sources as synchrotron radiation (synchro-Compton effect) or cosmic
micro-wave background radiation (CMBR): see \citet{Metal96},
\citet{Tetal00}, \citet{HK02} for recent reviews.\\   
Different acceleration mechanisms have been invoked so far to produce energetic
particles,  e.g. diffusive shock acceleration (DSA), second order Fermi
acceleration in a  magnetohydrodynamic wave turbulence (FII) \citep{BS87,
Hetal99}, shock drift  acceleration (SDA) \citep{BK90} and magnetic
reconnection (see for instance \citet{B96}, \citet{Betal01} and
references therein) \footnote{\citet{O00} and references therein
considered the  effect of tangential discontinuities in relativistic
jets.}. With some assumptions, all these mechanisms are able to accelerate
electrons up to TeV energies and are probably at work together in
extragalactic jets. Their combined effects have only been scarcely
discussed (see however \citet{CS92}, \citet{Metal99}), coupled shock
acceleration and spatial transport effects have been successfully applied
to hot spots by \citet{K62} (see also \citet{MK02} and reference
therein). Nevertheless, the resolution of the full convection-diffusion
equation governing the dynamical evolution  of the particle distribution
function does not usually lead to analytical solutions. The particle
transport and acceleration are closely connected to the local magneto-fluid
properties of the flow (fluid velocity fields, electro-magnetic fields,
turbulence). Recent  progress in computational modeling have associated
multidimensional fluid approaches (hydrodynamical (HD) or
magnetohydrodynamical codes (MHD)) with kinetic particle schemes
(\citet{Jetal02}, \citet{Jetal99}, \citet{Mietal99}). These
codes are able to describe the effects of shock and 
stochastic acceleration, adiabatic and radiative losses and the results are
used to produce synthetic radio, optical  and X-ray maps. The particle
transport is due to advection by the mean stream and turbulent flows. In
the ``Jones et al'' approach the shock acceleration process is treated using
Bohm prescription, i.e. the particle mean free path equals to the Larmor
radius). These above-mentionned treatments neglect spatial turbulent
transport and introduce spurious effects in the acceleration mechanism. This
leads to an overestimate of the particle acceleration efficiency in jets.\\  
In this work, we present a {\it new} method coupling kinetic theory and MHD
simulations in multi-dimensional turbulent flows. We applied the method to
the extragalactic {\it non-relativistic} or mildly relativistic (with a
bulk Lorentz factor $\Gamma_{\rm{jet}} < 2$) jets. Relativistic motions can
however be handled in case of non-relativistic shocks moving in a
relativistic jet flow pattern.\\ 
The paper covers from discussions about
turbulent transport and the coupling of kinetic schemes-MHD code to more
specific problems linked to jet physics. In Section \ref{Sacc} we review
the most important results concerning weak turbulence theory  and exposes
the effect of chaotic magnetic effects on the relativistic particle (RPs)
transport. Section \ref{Snum}  presents the system of stochastic
differential equations (SDEs) used to solved the diffusion-convection
equation of RPs. We examine the limits of the SDEs as regards to different
diffusion regimes and discuss their applicability to astrophysics.  Section
\ref{Sres} tests the ability of SDEs to describe accurately 
transport and acceleration of RPs in 2D versus known
analytical results. The MHD 
simulations of jets are  presented at this stage to investigate the problem
of shock acceleration.  Section \ref{SNI} treats RPs transport and
acceleration in complex flows configurations  as found in extragalactic
jets. We consider the problem of curved and non constant compression ratio
shocks. We derived analytical estimates on the expected particle  maximum
energy fixed by radiative losses or transversal escapes due chaotic
magnetic diffusivity  and MHD turbulence. We report our first results on
X-ray jets using MHD-SDE snapshots mixing  spatial transport, synchrotron
losses, strong single and multiple shock acceleration. We conclude in
Section \ref{Scon}.
\section{Acceleration and spatial transport}
\label{Sacc}
The accurate knowledge of transport coefficients is a key point to
probe the efficiency of the Fermi acceleration mechanisms as well as the
spatial transport of RPs in turbulent sources.  We assume a pre-existing
turbulent spectrum of plasma waves, retaining the Alfv\`en waves  efficient
to scatter off and accelerate charged particles. The particle trajectories
are random walks in space and energy, superimposed to the advection motion
induced by the  background flow, provided that the diffusion time is larger
than the coherence time of the  pitch angle cosine. If the
turbulence level, defined as the ratio of chaotic magnetic components  to total
one $\eta_T \ = \ <\delta B^2>/(<B^2+\delta B^2>)$ is much smaller than
unity, the spatial transport parallel to the mean magnetic field can be
described by the quasi-linear 
theory. Before discussing  any acceleration mechanism we shall recall the
main results of this theory and some of its non-linear  developments.

\subsection{Particle transport theories}
During its random walk on a timescale $\Delta t$ the position of the particle 
is changed by an amount $\Delta x_{\parallel}$ along the mean ordered 
magnetic field and by $\Delta x_{\perp}$ in the transverse direction.  The
ensemble average of both quantities vanishes, but the mean quadratic
deviations are non zero and define the parallel diffusion coefficient
$D_{\parallel} = <\Delta x_{\parallel}^2>/2 \Delta t$ and the perpendicular
diffusion coefficient $D_{\perp} = <\Delta x_{\perp}^2>/2 \Delta t$.\\
For a power-law turbulent spectrum $S(k) \propto \eta_T \ (k \
\lambda_{max})^{-\beta}$ completely defined by its turbulent level $\eta_T$,
spectral index $\beta$ and maximum turbulent scale $\lambda_{max}$,
the quasi-linear scattering frequency $\nu_s =<\Delta\cos^2(\theta)>/\Delta
t$ is \citep{Jok66}
\beq \nu_s = \eta_T \ \Omega_s \ |\mu|^{\beta-1}\ \tilde{\rho}^{\beta-1} \ .
\label{QSL0}
\eeq 
$\Omega_s$ is the synchrotron gyro-frequency $Z eB/ \gamma m_* c$ for a
particle of charge $Z e$, mass $m_*$ and Lorentz factor $\gamma$
and pitch-angle cosine $\mu = \cos\theta$. The Larmor radius 
$r_L = v/\Omega_s$ and the particle rigidity 
$\tilde{\rho} = 2 \pi r_L/\lambda_{max}$.\\ 
The scattering time $\tau_s$ is the coherence time of the pitch-angle cosine
and can be related to the pitch-angle frequency $\nu_s$ by $\tau_s\sim 1/\nu_s$ since 
the deflection of the pitch-angle typically occurs on one scattering time. 
The {\it quasi-linear diffusion} coefficients are
\begin{eqnarray}
D_{\parallel} &\ = \ &\frac{4}{5} \ \frac{v^2}{3} \ \tau_s \ \sim \frac{\lambda_{max} c}{3} \eta_T^{-1} 
\ \tilde{\rho}^{2-\beta} \ , 
\nonumber \\
D_{\perp} &\ = \ &\frac{v^2}{3} \ \frac{\tau_s}{1 + (\Omega_s \tau_s)^2} \ .
\label{QSL1}
\end{eqnarray}
Nevertheless, this simple approach does not take into account the
displacement of the guide-centers of particle trajectories. When magnetic
turbulence is occurring, the magnetic field lines are also diffusing, which
will amplify the transverse diffusion of particles following these magnetic
field lines \citep{Jok69}. Indeed including this effect in the diffusion
dynamics leads to a new transverse diffusion regime, namely the chaotic
transverse diffusion \citep{Rech78,Rax92}.         The work done by
\citet{Cass02} presents extensive Monte-Carlo simulations of charged
particles in a magnetic field composed of a regular and a turbulent part,
calculated assuming power-law spectra of index $\beta$ (as in Kolmogorov or
Kraichnan theories). The authors present, using  averaged spatial
displacements over time intervals, the behavior of the spatial diffusion
coefficients as a function of the particles energies as well as turbulence
level $\eta_T$. The diffusion coefficient along the mean magnetic field
displays energetic dependence similar to the quasi-linear theory but on any
turbulence level. On the other hand, the diffusion coefficient transverse to
the mean magnetic field is clearly in disagreement with neo-classical
prediction (see Eq.~\ref{QSL1}). The chaotic transverse diffusion  regime
is occurring when the turbulence level is large but can
probably be extended to lower turbulent levels, as first imagined by
\citet{Rech78}. In \citet{Cass02} this regime was observed for all
turbulence levels down to $\eta_T=0.03$. The resulting transverse
coefficient is reduced to 
$D_{\perp}\propto D_{\parallel}$  with a proportionality factor only
depending on 
the turbulence level, namely
\begin{eqnarray}
D_{\parallel} &\propto&
\frac{c\lambda_{max}}{\eta_T} \ \tilde{\rho}^{2-\beta} \ ,
\nonumber \\
D_{\perp} &\propto&
\eta_T^{1.3}c\lambda_{max} \ \tilde{\rho}^{2-\beta} \ .
\label{CDi1}
\end{eqnarray}
In this paper we will use the above prescription as, unless very low
$\eta_T$, the chaotic diffusion always dominates.  

\subsection{Acceleration processes}
\label{SSDA}
In a diffusive shock \footnote{The shock drift acceleration mechanism has
been applied to electron acceleration in extragalactic radio sources by
\citet{AV93} and references therein. This effect will not be considered in
the simulations and is not further discussed} particles able to resonate
with wave turbulence, undertake a pitch-angle scattering back and forth
across the shock front gaining energy. The finite extension of the
diffusive zone implies some escapes  in the downstream flow. The stationary
solution for a non-relativistic shock can be written as $f(p) \propto
p^{-(3+\tau_{acc}/\tau_{esc})}$. In a strong shock the acceleration
timescale $\tau_{acc}$ exactly balances the  particle escape time scale
$\tau_{esc}$ \citep{D83}. The acceleration timescale, for a parallel shock
is $\tau_{accDSA} = 3/(r-1) \ t_r$, where $r=u_u/u_d$ is the shock
compression ratio ($u_u$ and $u_d$ are respectively upstream and downstream
velocities of the fluid in the shock frame) and $t_r = (c/u_d)^2 \tau_s$ is
the downstream  particle residence time.\\ The MHD turbulence, especially
the Alfv\`en turbulence, mainly provokes a diffusion of the particle
pitch-angle.  But the weak electric field of the waves $\delta E/\delta B
\equiv V_a/c$ also accelerates particles. The momentum diffusion is of
second order in terms of Fokker-Planck description and the acceleration
timescale is $\tau_{accFII} = (c/V_a)^2 \ \tau_s$.  Note that even if the
stochastic acceleration is a second order process,  $\tau_{accFII}$ may be
of the same order as $\tau_{accSDA}$ in low (sub-alfvenic) velocity flows
or high Alfv\`en speed media as remarked by \citet{Hetal99}.\\ In radio
jets (see \citet{F85} and \citet{F98} for reviews of jet properties)
equiparition between magnetic fields and non-thermal, thermal plasmas lead
to typical magnetic fields $B \sim 10^{-5/-4} \ \rm{Gauss}$, thermal proton
density $n_p \sim 10^{-2/-5} \ \rm{cm^{-3}}$ and thus to Alfv\`en speeds
$V_a/c$ between $7\times 10^{-4}-0.2$. In light and magnetized jets, the
second order  Fermi process can be faster than diffusive shock
acceleration. We decided to postpone the effect of second order Fermi
acceleration to a future work. In this first step, we mostly aim to
disentangle the  diffusive shock acceleration process, the turbulent
spatial transport and radiative losses effects shaping  the particle
distribution. We will therefore only consider super-Alfvenic flows
hereafter.

\section{Numerical framework}
\label{Snum}
In this section, we present the multidimensional stochastic differential equations system
equivalent to the diffusion-convection equation of RPs.\footnote{\citet{SA01} have investigated
the coupling between 2D Hydrodynamical code and SDEs adapted to the nonthermal X-ray emission from 
supernova remnants.}
\subsection{Stochastic differential equations}
The SDEs are an equivalent formulation of the Fokker-Planck equations
describing the evolution of the distribution function of a particle
population. It has been shown by \citet{Ito51} that the
distribution function $f$ obeying Fokker-Planck equation as
\begin{eqnarray}
\frac{\p f}{\p t} =-\sum_{i=1}^{N}\frac{\p}{\p
X_i}\left(A_i(t,\XX)f(t,\XX)\right) + \nonumber\\
\frac{1}{2}\sum_{i=1}^{N}\sum_{j=1}^N\frac{\p^2}{\p X_i\p
X_j}\left(\sum_{k=1}^NB_{ik}(t,\XX)B_{kj}^T(t,\XX)f(t,\XX)\right)
\label{SDE1}
\end{eqnarray}
at a point $\XX$ of phase space of dimension $N$, can also be described as a set 
of SDEs of the form \citep{Krul94}
\begin{eqnarray} 
\frac{d\XX_{t,i}}{dt}=A_i(t,\XX_t)
+\sum_{j=1}^NB_{ij}(t,\XX_t)\frac{dW_{t,j}}{dt} \ ,\nonumber\\
&& i=1,..,N
\label{SDE2}
\end{eqnarray}
\noindent where the $W_{t,j}$ are Wiener processes satisfying $<W>=W_o$ and
$<(W-W_o)^2>= t-t_o$ ($W_o$ is the value of $W$ at $t_o$). The diffusion
process described by Fokker-Planck equations can be similarly taken into
account if $dW_i/dt=\xi_i$ is a random variable with a Gaussian
conditional probability
such as 
\beq
p(t,\xi|t_o,\xi_o)=\frac{1}{\sqrt{2\pi(t-t_o)}}\exp\left(-\frac{(\xi-\xi_o)^2}{2(t-t_o)}\right)\ .   
\label{SDE3}
\eeq   
The Fokker-Planck equation governing this population will be \citep{Skil75}
\begin{eqnarray}  
\frac{\p f}{\p t}&=&-(\UU\cdot{\bf \nabla})f
 +\frac{1}{3}({\bf\nabla}\cdot\UU)p\frac{\p f}{\p p}+\nabla_i.(D_{ij}\nabla_j f)\nonumber\\
 &+&\frac{1}{p^2}\frac{\p}{\p
p}\left(D_{pp}p^2\frac{\p f}{\p p}+a_{syn}p^4f\right) \ ,
\label{SDE4}
\end{eqnarray}
\noindent where $D_{ij}$ is the spatial diffusion tensor and
$D_{pp}$ describes energy diffusion in momentum space. The term $a_{syn}$
stands for synchrotron losses of the electrons. Its expression is
\beq
a_{syn} = \frac{\sigma_TB^2}{6\pi m_e^2c^2} \ ,
\eeq
\noindent where $\sigma_T$ is the Thomson cross-section. This term can easily 
be modified to account for Inverse Compton losses. In
term of the variable $F=Rp^2 f$, these equations can be written in cylindrical
symmetry (R varies along the jet radius and Z along the axial direction)
\begin{eqnarray}
\frac{\p F}{\p t}= &-&\frac{\p}{\p R}\left(F\left\{U_R+\frac{\p D_{RR}}{\p R}
+\frac{D_{RR}}{R}\right\}\right)\nonumber\\
& -& \frac{\p}{\p Z}\left(F\left\{U_Z+\frac{\p
D_{ZZ}}{\p Z}\right\}\right)\nonumber \\
&-&\frac{\p}{\p p}\left(F\left\{-\frac{p}{3}\nabla\cdot\UU+\frac{1}{p^2}\frac{\p p^2
D_{pp}}{\p p}-a_{syn}p^2\right\}\right)\nonumber\\
&+&\frac{\p^2}{\p R^2}(FD_{RR})+\frac{\p^2}{\p Z^2}(FD_{ZZ})+\frac{\p^2}{\p
p^2}(FD_{pp})
\label{SDE5}
\end{eqnarray}
\noindent Note that this rewriting of the Fokker-Planck equation is valid
only if $Rp>0$. Assuming that the diffusion tensor is diagonal, it
is straightforward to get the SDEs coefficients. These equations can then be
written as  
\begin{eqnarray}
\frac{dR}{dt}&=&U_R+\frac{\p D_{RR}}{\p
R}+\frac{D_{RR}}{R}+\frac{dW_R}{dt}\sqrt{2D_{RR}} \ , \\
\frac{dZ}{dt}&=&U_Z+\frac{\p D_{ZZ}}{\p
Z}+\frac{dW_Z}{dt}\sqrt{2D_{ZZ}} \ , \\
\frac{dp}{dt}&=&-\frac{p}{3}\nabla\cdot\UU+\frac{1}{p^2}\frac{\p
p^2D_{pp}}{\p p}-a_{syn}p^2 \nonumber \\
&+&\frac{dW_P}{dt}\sqrt{2D_{pp}} \ .
\label{SDE6}
\end{eqnarray}
\noindent where $U_{R/Z}$ stand for the radial and axial component of fluid
velocity field. The $W$ are stochastic variables described previously. They are computed
using a Monte-Carlo subroutine giving a random value $\xi$ with zero mean and 
unit variance so that we can build the trajectory of one 
particle in phase space from time $t_{k}$ to $t_{k+1}=t_{k}+\D t$ \citep{Marc99}
\begin{eqnarray}
R_{k+1}=&R_k&+\left(U_R+\frac{1}{R}\frac{\p RD_{RR}}{\p
R}\right)_{k}\D t\nonumber\\
&+&\xi_R\sqrt{2D_{RR}\D t} \ . \\
Z_{k+1}=&Z_k&+\left(U_Z+\frac{\p D_{ZZ}}{\p
Z}\right)_{k}\D t +\xi_Z\sqrt{2D_{ZZ}\D t} \ . \\
p_{k+1}=&p_k&+\left(-\frac{p}{3}\nabla\cdot\UU+\frac{1}{p^2}\frac{\p
p^2D_{pp}}{\p p}-a_{syn}p^2\right)_k\D t\nonumber\\
&+&\xi_p\sqrt{2D_{pp}\D t}\ .
\label{SDE7}
\end{eqnarray}    
\noindent It is noteworthy that these algorithms derived from SDEs are
only valid if the particles are not at the exact location of the jet axis,
otherwise an unphysical singularity would occur. The coupling between the
SDEs and a macroscopic simulations clearly appears here. The macroscopic
simulation of the jet, using magnetohydrodynamics, would give the
divergence of the flow velocity as well as the strength and the orientation
of the magnetic field at the location of the particle. Indeed, as shown in
the last paragraph, the spatial diffusion of particles is mainly driven by
the microscopic one, namely by the magnetic turbulence. Since the work of
\citet{Cass02}, the behavior of the diffusion coefficients both along and
transverse to the mean magnetic field are better known. They depend on the
strength of the mean magnetic field, on the particle energy, and on the
level of the turbulence $\eta_T$. Once the diffusion  coefficients are
known, the distribution function is calculated at a time t at the shock
front by summing the particles crossing the shock between t and
t+$\rm{\Delta t}$. The  distribution function can in princinple be
calculated everywhere if statistics are good enough. We typically used
$5\times 10^5-10^6$ particles per run.

\subsection{Constrains on SDE schemes}

\subsubsection{Scale ordering}
\label{Sordering}
Particles gain energy in any compression in a flow. A compression is
considered as a shock if it occurs on a scale much smaller than the test
RPs mean free path. The acceleration rate of a particle with momentum
$p$ through the first-order Fermi process is given by the divergence term
in Eq.(\ref{SDE6}), e.g. $<\frac{dp}{dt}> = - \frac{p}{3} \
\nabla.\bf{v}$. The schemes used in the present work are explicit
\citep{Krul94}, i.e. the divergence is evaluated at the starting position
$x(t_k)$.  
Implicit schemes \citep{Marc99} use the velocity field at the final position 
$x(t_{k+1})$ to compute the divergence as
$(u(t_{k+1})-u(t_k))/(x(t_{k+1})-x(t_k))$.\\ 
The particle walk can be decomposed into an advective and a diffusive step evaluated at $t_k$
and incremented to the values $R(t),Z(t),p(t)$ to obtain the new values at
$t_{k+1}$. As demonstrated by \cite{SG89} and \cite{KP92} it is possible to 
expand the It\^o schemes into Taylor series
to include terms of higher order in $\Delta t$ and in turn to improve the
accuracy of the algorithms. However, both because higher order schemes need
to store more data concerning the fluid (higher order derivatives) and the
schemes have proved to accurately compute the shock problem in 1D, we only
use explicit Euler (first order) schemes in the following simulations.\\
Hydrodynamical codes usually smear out shocks over a given number of grid
cells because of (numerical) viscosity. The shock thickness in 2D is then a
vector whose components are $\Delta X_{\rm{shock}} =((\alpha_r \ \Delta
R),(\alpha_z \ \Delta Z))$, where $(\Delta R,\Delta Z)$ describes one grid
cell  and the coefficients $(\alpha_r,\alpha_z)$ are typically of the order
of a few. We can construct, using the same algebra, an advective $\Delta
X_{\rm{adv}}$ and a diffusive $\Delta X_{\rm{diff}}$ vectors steps from
equations (10-\ref{SDE6}). \cite{Krul94} have found that a SDE scheme can
correctly calculate the effects of 1D shock acceleration if the different
spatial scales of the problem satisfy the following inequality \beq \Delta
X_{\rm{adv}} \ll X_{\rm{shock}} < \Delta X_{\rm{diff}}\ .
\label{Explicit}
\eeq
In 2D this inequality must be fulfilled by each of the vector components, e.g.
\begin{eqnarray}
\Delta R_{\rm{adv}} &\ll \alpha_r \Delta R < &\Delta R_{\rm{diff}} \ , \nonumber \\
\Delta Z_{\rm{adv}} &\ll \alpha_z \Delta Z < &\Delta Z_{\rm{diff}} \ .
\label{Explicit2d}
\end{eqnarray}
These are the {\it 2D explicit schemes} conditions for the computation of
shock acceleration. 
The two previous inequalities impose constrains on both the simulation
timescale $\Delta t$, and the diffusion coefficients $D_{RR}$ and $D_{ZZ}$.

\subsubsection{Minimum diffusion coefficients}
The finite shock thickness results in a lower limit on the diffusion
coefficient. The condition $\Delta X_{\rm{adv}}  \ll \Delta X_{\rm{diff}}$
implies a maximum value for the time step $\Delta t_{\rm SDE}$ that  can be
used in the SDE method given a fluid velocity $\bf{u}$ (omitting for
clarity the terms including the derivatives of the diffusion coefficients)
\beq \Delta t_{\rm max} \ = \ \Delta X_{\rm shock} /|u|  .
\label{hydro}
\eeq
In 2D we shall take the {\it minimum} $\Delta t_{\rm{max}}$ thus derived
from (\ref{Explicit2d}).\\ 
Inserting this time step into the second part of the
restriction, $\Delta X_{\rm{adv}}  
\ll \Delta X_{\rm{diff}}$ yields a {\it minimum} value for the diffusion 
coefficient:
\beq
\label{Minimum}
D_{\rm min} \ = \ \frac{1}{2}|u|\ X_{\rm shock}  \ .  
\eeq 
If the diffusion coefficient depends on momentum, this condition implies
that there is a 
limit on the range of momenta that can be simulated. The fact that the
hydrodynamics sets a  limit on the range of momenta may be inconvenient in
certain applications. One can in principle circumvent this problem by using
adaptive mesh refinement \citep{Berg86,leveque98}.  This method
increases the grid-resolution in those regions where more resolution  is
needed, for instance around shocks.  The method is more appropriate than
increasing the resolution over the whole grid. \\ 
Another possibility would
be to sharpen artificially shock fronts or to use an implicit SDE scheme. 
This approach can be useful in one dimension, but fails in
2D or 3D when the geometry of shock fronts becomes very complicated, for
instance due to corrugational instabilities.\\

\subsubsection{Comparisons with other kinetic schemes}
As already emphasized in the introduction, up to now, few works have
investigated  
the coupling of HD or MHD codes and kinetic transport schemes (mainly
adapted to the 
jet problem). The simulations performed by T. Jones and co-workers
(\citet{Jetal02}, 
\citet{Tetal01}, \citet{Jetal99}) present 2 and 3-dimensional synthetic
MHD-kinetic radio jets including diffusive acceleration at shocks as well
as radiative and adiabatic cooling.  They represent a great improvement
compared to previous simulation where the radio emissivity was scaled to
local gas density. The particle transport is treated solving a
time-dependent diffusion-convection equation. The authors distinguished two
different jet regions: the smooth flows regions between two sharp shock
fronts where the leading transport process is the convection by the
magneto-fluid and the shock region where the Fermi first order takes
over. This method can account for stochastic acceleration in energy, but the
process have not been included in the published works.  The previous
distinction relies on the assumption that the electron diffusion length  is
smaller than the dynamical length as it is the case for Bohm diffusion (see
discussion  in section \ref{Sordering}). However, Bohm diffusion is a very 
peculiar regime  appearing for a restricted rigidity ranges (see
\cite{Cass02} and Eq.~\ref{CDi1}).  The magnetic chaos may even
completely avoid it. It appears then essential to encompass  diffusive
spatial transport within MHD simulations.\\ 
\citet{Mietal99} computed the spatial and energy time transport of Lagrangian 
cells in turbulent flows generated by Kelvin-Helmholtz (KH) instability. The 
energetic spectrum of a peculiar cell is the solution of a spatially averaged 
diffusion-convection equation \citep{K62}. This approach accounts for the 
effect of fluid  turbulence on the particle transport but suffers from 
the low number of Lagrangian cells used to explore the jet medium. 
The SDE method has the advantage to increase considerably the statistics 
and to allow the construction of radiative maps. As the particles are embedded 
in the magnetized jet both macroscopic (fluid) and microscopic (MHD waves, 
magnetic field wandering) turbulent transport are naturally included in the simulations.

\section{Testing coupling between MHD and SDEs}
\label{Sres}
So far, particle energy spectra produced by SDEs were calculated using one
dimensional prescribed velocity profiles (see however \citet{SA01}). 
The prescriptions described plane shocks as a velocity discontinuity 
or as a smooth velocity transition. The aim of this section is to present energetic
spectra arising from {\it shocks generated by macroscopic numerical
code}. 
The tests will increasingly be more complex including different effects
entering in particle transport and acceleration in extragalactic jets.\\
In the first subsection, we present very elementary tests devoted
to control the accuracy of the particle transport by SDEs in a cylindrical
framework. In the second part, we move to jet physics and present
the MHD jet simulations and discuss the results of particle
acceleration in near-plane shocks produced by Kelvin-Helmholtz
instabilities.
\subsection{Testing 2D cylindrical diffusive transport}
\noindent Before proceeding to any simulations where both MHD and SDE are
coupled, we  have tested the realness of our description of the spatial
transport of relativistic particles. Testing SDEs has already been
addressed by \citet{Marc99} and references therein but only in a
one-dimensional framework. They successfully described the particles
acceleration by thin shocks as well as the synchrotron emission occurring 
in the case of relativistic electrons. Here, we have added a 
second spatial SDE, for the radial transport, where extra-terms appear 
because of the cylindrical symmetry. One way to test the 2D transport is to
compute the 
confinement time of a particle set in the simple case of {\it a uniform jet
with uniform diffusion coefficient} $D_{RR}$ and $D_{ZZ}$.
Let assume we have a set of $N$ particles at the jet axis at
$t=0$. The diffusion process will tend to dilute this population in space
and after some time, most of the particles will leave the plasma
column. Indeed, the average position will be the initial position but the
spatial variance of these particles at time $t$ will be $\sqrt{2Dt}$. For
the specific problem of a cylindrical jet, let consider a cross section in
the Cartesian $X$ and $Y$ directions while $Z$ is along the jet axis. The
set of particles will stop to be confined once 
\beq R^2_{jet} \leq 2D_{XX}
t + 2D_{YY} t \ ,
\label{test1}
\eeq
where $R_{jet}$ is the jet radius and the diffusion coefficients $D_{XX}$ and $D_{YY}$ 
can be related to $D_{RR}$ by 
\begin{eqnarray}
D_{RR} &=& \frac{<\Delta R^2>}{2\Delta t} = \left<\frac{(X\Delta X+Y\Delta
Y)^2}{R^2}\right>\frac{1}{2\Delta t}\nonumber\\
&=& D_{XX} = D_{YY}\ .
\label{test2}
\end{eqnarray} 
\noindent In this relation, $X$ and $Y$ are two uncorrelated variables
($<\Delta X\Delta Y>=0$). It is then easy to see that the confinement time
of a set of particles inside a jet is
\beq
T_{cf} = \frac{R^2_{jet}}{4D_{RR}}
\label{test3}
\eeq  
\noindent when one consider an infinitely long jet (no particle escape in
the $Z$ direction). 
\begin{figure}[t]
\resizebox{\hsize}{!}{\includegraphics{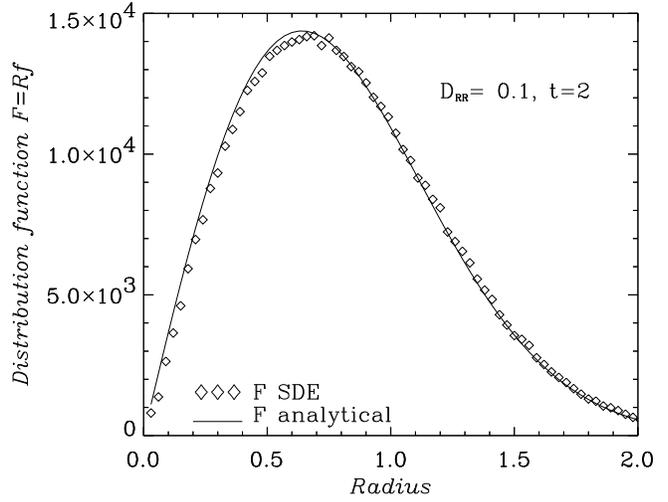}}
\caption{Plot of the distribution function $F=Rf$ modelized by SDE in the
case of a uniform spatial diffusion, for a fixed $Z$ versus the radial coordinate in
jet radius unit. The solid curve is
the analytical solution obtained from Fokker-Planck equation Eq.(\ref{SDE4}) which is in good
agreement with computations using SDEs.}
\label{f1}
\end{figure}
\noindent We have performed a series of calculations dealing with one million
particles injected near the jet axis with different values of the radial
diffusion coefficient. We have set a time step of $\Delta t=5.10^{-3}$ and
integrated the particles trajectories using the numerical scheme
Eq.(\ref{SDE7}). When a particle has reached the jet surface ($R=R_{jet}$),
we stop the integration and note its confinement time. Once all particles
have reached the jet surface, we calculate the average value of the
confinement time. In Tab.~\ref{tab2}, we present
the result of the different computations. The good agreement between the
numerical and the estimated confinement times is a clue indicating that the
spatial transport of the particles in the jet is well treated as far as the
time step is small enough to mimic the Brownian motion of particles.
\begin{table}[t]
\begin{center}
\begin{tabular}{c||cc}
$D_{RR}$ & $R^2_{jet}/4D_{RR}$ & $T_{cf}$ \\ \hline
$0.0125 $ & $20$ & $19.96$ \\
$0.025 $   &  $10$  & $9.94$\\
$0.075$ & $10/3$ & $3.25$ \\
$0.15$ & $5/3$ & 1.47 
\end{tabular}
\end{center}
\label{tab2}
\caption{Computations of confinement time $T_{cf}$ for different
diffusion coefficient values and theoretical value of this confinement
time. Note that the agreement is good as far as the confinement time is
large. Indeed, the time step $\Delta t=5\times10^{-3}$ to compute them is the same
for the three runs which leads to different ratio $T_{cf}/\Delta
t$. If this ratio is too small, the time step is not appropriate to nicely
modelize the particle transport.}
\end{table}
\noindent Another way to test SDEs in this problem is to look at the
distribution function of these particles since the analytical solution to
the diffusion with uniform coefficients is known. The Fokker-Planck
equation, in the case of a uniform spatial diffusion without any energetic
gains or losses, is 
\beq
\frac{\p f}{\p t}= \frac{D_{RR}}{R}\frac{\p}{\p R}\left(R\frac{\p f}{\p
R}\right) + D_{ZZ}\frac{\p^2 f}{\p Z^2}
\label{test4}
\eeq
\noindent The radial dependence of f arising
from this equation is, for an initial set of particles located at the jet
axis,  
\beq
f(R,Z,t) \propto  \frac{1}{4D_{RR}t}\exp\left(-\frac{R^2}{4D_{RR}t}\right)\ .
\label{test5}
\eeq  On Fig.~\ref{f1} we plot the distribution function $F=Rf$ obtained 
for a set of $5\times10^5$ particles located initially
very close to the jet axis. The plot is done at a given time $t=2$ and with
$D_{RR}=D_{\perp}=0.1$. The symbols represent the numerical values obtained
using SDEs while the solid line represents the analytical solution from
Eq.(\ref{test5}). The good agreement between the two curves is a direct
confirmation that the transport of particles is well modelized by SDEs.

\subsection{MHD simulations of extragalactic jets}
\label{mhdsimu}
\noindent In order to describe the evolution of the jet structure, we have
employed the Versatile Advection Code  (VAC, see \citet{Toth96a} and {\tt
http://www.phys.uu.nl/}$\sim${\tt toth}).  We solve the set of MHD
equations under the assumption of a cylindrical symmetry. The initial
conditions described above are time advanced using the conservative, second
order accurate Total Variation Diminishing Lax-Friedrich scheme \citep{Toth96b}
with minmod limiting applied on the primitive variables. We use a
dimensionally unsplit, explicit predictor-corrector time marching. We
enforce the divergence of the magnetic field to be zero by applying a
projection scheme prior to every time step \citep{Bruc80}.
\subsubsection{MHD equations}
We assume the jet to be described by ideal MHD in an axisymmetric
framework. This assumption of no resistivity $\nu_m$ has  consequences on the 
particle acceleration since the Ohm law states the electric field as ${\bf
E}=-\UU\times\BB$. This electric field will vanish in the fluid rest frame
so that no first-order Fermi acceleration can be achieved by ${\bf E}$. In the
case of a resistive plasma, the electric field (${\bf E}=\BB\times\UU +
\nu_m{\bf J}$, ${\bf J}$ density current) cannot vanish by a frame
transformation and a first-order Fermi acceleration will occur. 
\noindent In order to capture the dynamics of shocks, the VAC code has been
designed to solve MHD equations in a conservative form, namely to insure conservation of
mass, momentum and energy. The mass conservation is 
\beq
\frac{\p \rho}{\p t} + \nabla\cdot(\rho \UU_p) = 0 \ ,
\label{mhd1}
\eeq
\noindent where $\rho$ is the density and $\UU_p$ is the \pol \ component of
the velocity. The momentum conservation has to deal with both thermal
pressure gradient and MHD Lorentz force, namely
\beq
\frac{\p \rho\UU}{\p t} + \nabla\cdot\left(\UU\rho\UU -\frac{\BB\BB}{\mu_o}\right) +
\nabla(\frac{B^2}{2\mu_o}+ P) = 0 \ ,
\label{mhd2}
\eeq  
\noindent where $P$ stands for thermal plasma pressure. The induction
equation for the magnetic field is
\beq 
\frac{\p \BB}{\p t}= - \nabla\cdot(\UU\BB - \BB\UU) \ ,
\label{mhd3}
\eeq
The last equation deals with the energy conservation. The total energy  
\beq
e=\frac{\rho\UU^2}{2}+\frac{\BB^2}{2\mu_o}+\frac{P}{\Gamma-1} \ ,
\label{mhd4}
\eeq 
\noindent where $\Gamma=C_P/C_V=5/3$ is the specific heat ratio, is governed by  
\beq  
\frac{\p e}{\p
t}+\nabla\cdot\left(\UU
e-\frac{\BB\BB}{\mu_o}\cdot\UU+\UU\left[P+\frac{\BB^2}{2\mu_o}\right]\right)=  0 \ .
\label{mhd5}
\eeq
In order to close the system of MHD equations, we assume the plasma as
perfect gas. Thermal pressure is then related to mass density and temperature as
\beq
P = \frac{\Re}{\mu_p}\rho T
\label{mhd6}
\eeq
\noindent where $\Re$ is the perfect gas constant and $\mu_p$ the plasma mean
molecular weight.\\   
By definition, these simulations are not able to describe microscopic
turbulence since MHD is a description of the phenomena occurring in a
magnetized plasma over large distance (typically larger than the Debye
distance to insure electric charge quasi-neutrality). So, in the case of
diffusion coefficients involving magnetic turbulence, we shall have to
assume the turbulence level $\eta_T$. 
\subsubsection{Initial conditions and boundaries}
 We consider an initial configuration of the structure such as the jet is a
plasma column confined by magnetic field and with an axial flow. We add to
this equilibrium a radial velocity perturbation that will destabilize the
flow to create Kelvin-Helmholtz instabilities. The radial balance of the jet
is provided by the opposite actions of the thermal pressure and magnetic force
\begin{eqnarray}
B_Z(R,Z,t=0)&=&  1 \ , \nonumber\\
B_R(R,Z,t=0)&=& 0 \ , \nonumber\\
B_{\theta}(R,Z,t=0)&=& -\frac{(R/R_c)}{1 + (R/R_c)^2}\ , \nonumber\\
P(R,Z,t=0)&=& \left[\frac{1}{(1 + (R/R_c)^2)^2}+ \beta_p -1\right]
\label{IS3} \ .
\end{eqnarray}
\noindent where $R_c$ is a parameter controlling the location of the
maximum of $B_{\theta}$ and $\beta_p=2\mu_oP_o/B^2_o$  is the ratio of
thermal to magnetic pressure at the jet axis. All magnetic field 
components are here expressed in $B_o$ units (see next for a definition).\\
The FRI jets are partially collimated
flows where some instabilities seem to perturb the structure of the
jet. Thus we will assume in our simulation that the thermal pressure is not
negligeable in the jet and that the flow is prone to axisymmetric
Kelvin-Helmholtz instabilities. Thus we will assume values of $\beta_p$
larger than unity. The sonic Mach number is implemented as 
\beq
\frac{u_Z(R,Z,t=0)}{C_o}=M_s=\frac{M_o}{\cosh((R/R_o)^8)}      
\label{IS2}
\eeq
\noindent where $C_o$ is the sound speed at the jet
axis. This sound speed can be related to jet velocity $U_{jet}$ by the
parameter $M_o=U_{jet}/C_o$. This
parameter is chosen to be larger than one, as the jet is expected to
be super-fastmagnetosonic. In our simulations, we have $M_o=10$ and 
$u_{\theta}=0$. The perturbation that can provoke
KH instabilities must have a  velocity component perpendicular to the flow
with a wave vector parallel to the flow (e.g. \citet{Bodo94}). We have then
considered a radial velocity perturbation as
\beq 
\frac{u_R(R,Z,t=0)}{C_o}=\delta M_o
\frac{\sum_{k=1}^{n_z}\sin(kZ2\pi/L_o)}{\exp\left(5(Z-R_o)^2\right)}   
\label{IS4}
\eeq
\noindent where $\delta M_o$ is smaller than unity in order to create a
sub-sonic perturbation and $L_o$ is the vertical length of the box. The
density of the plasma is set as 
\beq
\rho(R,Z,t=0)= \left(\frac{0.8}{\cosh((R/R_o)^8)}+0.2\right)
\label{IS1}
\eeq
where $\rho_o=\Gamma\beta_pB^2_o/2\mu_oC_o^2$ is the density at the jet axis.\\

\noindent \underline{{\it Physical quantities normalization}}:\\
Lengths are normalized to the jet radius $R_o$ at the initial
stage. The magnetic field physical value is given by $B_o$ while velocities
are scaled using sound speed $C_o=U_{jet}/M_o$ intimately related the
observed jet velocity. Once physical values are assigned to the
above-mentionned quantities, it is straightforward to obtain all the other
ones. The dynamical timescale of the structure is expressed as 
\beq
\tau_o = \frac{R_o}{C_o} = 3.25\times10^4 yr \ 
M_o\left(\frac{R_o}{100pc}\right)\left(\frac{U_{jet}}{3000km/s}\right)^{-1}\\
\label{IS5}
\eeq
Note that for the single MHD simulations, the evolution of the structure
does not depend on these physical quantities but only on the parameters
$\beta_p,M_o,R_c,\delta M_o$. Nevertheless, for each MHD-SDE computation,
the physical values are injected into SDE equations to derive the velocity
divergence and the synchrotron losses.\\  

The MHD simulation are performed using rectangular mesh of size
$104\times 204$ cells, with two cells on each side devoted to boundary
conditions. The left side of the grid ($R=0$) is treated as the jet axis,
namely assuming symmetric or antisymmetric boundaries conditions for the
set of quantities (density, momentum, magnetic field and internal
energy). The right side of the box is at $R=4R_{o}$ and is consistent with
free boundary: a zero gradient is set for all quantities. For the bottom
and upper boundaries (respectively at $Z=0$ and $Z=8R_{o}$, we prescribe
periodic conditions for all quantities, so that when a particle reaches one
of these regions, 
it can be re-injected from the opposite region without facing
artificial discontinuous physical quantities.    
\begin{figure}
\resizebox{\hsize}{!}{\includegraphics{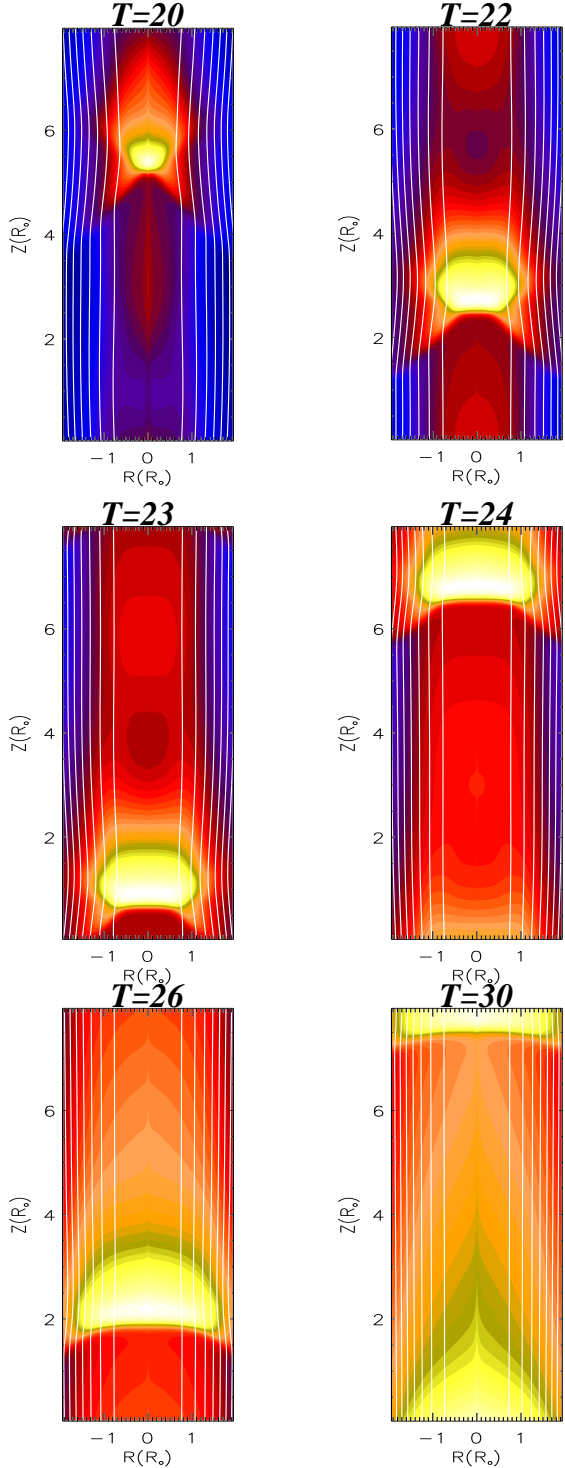}}
\caption{ Temporal evolution (in $\tau_o$ unit) of a typical internal
shock occurring within the jet. The grey-scales represent density levels
(dark for low density and 
white for high density) while solid lines stand for \pol \ magnetic field
lines. The parameters of this simulations are $\beta_p=10$, $M_o=10$,
$R_c=1$ and $\delta M_o=0.1$. This shock arises from a initial setup prone to
axisymmetric MHD 
Kelvin-Helmholtz instabilities. In the early stage of the shock evolution,
the shock front displays a bow-shock shape but as the simulation goes on,
the shock front evolves toward a front shock. }
\label{f2}
\end{figure} 
\subsubsection{Inner-jet shock evolution}
\label{Injet}
Kelvin-Helmholtz instability is believed to be one of the source of flow
perturbation in astrophysical jets. The evolution of this mechanism has been
widely investigated either in a hydrodynamical framework
(e.g. \citet{Mico00} and reference therein) or more recently using MHD
framework (see \cite{Baty02} and reference therein). The growth and
formation of shock as well as vortices in the jet core depend on the nature
of the jet (magnetized or not) and on the magnetic field strength
\citep{Mala96,Fran96,Jone97,Kepp99}. In the particular case of
axisymmetric jets, it has been shown that the presence of a weak magnetic
field significantly modifies the evolution of the inner structures of
vortices.\\ 
We present on Fig.~\ref{f2} the temporal evolution of a
typical inner-jet shock obtained from our computations. After the linear
growth of the instability (up to $t=19\tau_o$), the structure exhibits a curved
front shock inclined with respect to the jet axis. In the frame of the
shock, the flow is upstream super-fastmagnetosonic, and downstream
sub-fastmagnetosonic. On both side of the shock, the plasma flow remains
superalfv\`enic. This shock configuration is consistent with a super-fast
shock. Rankine-Hugoniot relations show that, at a fast-shock front, the
magnetic field component parallel to the shock front is larger in the
downstream medium than in the upstream one \citep{Frai91}. In the present
axisymmetric simulations, the bending of the \pol \ magnetic field lines
occurring at the shock front creates a locally strong Lorentz force that
tends to push the structure out of the jet. As seen on the following
snapshots of Fig.\ref{f2}, the shock front rapidly evolves toward a plane
shape.  This quasi-plane shock structure remains stable for several time 
units before being diluted.

\subsubsection{Macroscopic quantities}
\label{macroc}
The SDEs coupled with the MHD code provide approximate solutions of the
Fokker-Planck equation using macroscopic quantities calculated by the MHD
code.  Indeed, flow velocity and magnetic field enter the kinetic transport
equation  and there is no way to treat realistic case in astrophysical
environments but  to model them from macroscopic multi-dimensional
simulations. Nevertheless,  one difficulty remains since MHD (or HD)
simulations only give these macroscopic  quantities values at discrete
location, namely at each cells composing the numerical mesh. Hence, these
values are interpolated from the grid everywhere in the computational
domain. If the domain we are considering is well-resolved (large number of
cells in each direction), a simple tri-linear interpolation is sufficient
to capture the local variation of macroscopic quantities. When shocks are
occurring, the sharp transition in velocity amplitude is more difficult to
evaluate because shocks are typically only described by few
cells. Thus, the calculus of velocity divergence must be done accurately. We
adopt the following procedure to calculate it: shocks are characterized by
very negative divergence so at each cells $(i,j)$ we look for the most
negative result from three methods
\begin{eqnarray}
&&\nabla\cdot\UU(i,j) = min\left(\mp\frac{u_Z(i,j)-u_Z(i,j\pm 1)}{|Z(j)-Z(j\pm
1)|},\right.\nonumber\\
&&\left.\frac{u_Z(i,j+1)-u_Z(i,j-1)}{Z(j+1)-Z(j-1)}\right)\nonumber \\
&+& \min\left(\mp\frac{R(i)u_R(i,j)-R(i\pm 1)u_R(i\pm
1,j)}{R(i)(R(i)-R(i\pm 1))},\right.\nonumber\\
&&\left.\frac{R(i+1)u_R(i+1,j)-R(i-1)u_R(i-1,j)}{R(i)(R(i+1)-R(i-1,j))}\right)
\label{div}
\end{eqnarray}
\noindent This approach ensures that the sharp velocity variation occurring
within a shock is well described and that no artificial smoothing is
created in the extrapolation of flow velocity divergence. At last, note
that the location of the most negative $\nabla\cdot\UU$ corresponds to the
shock location. The measurement of spectra at shock front will then be done
by looking at particles characteristics passing through this location.

\begin{figure}[t]
\resizebox{\hsize}{!}{\includegraphics{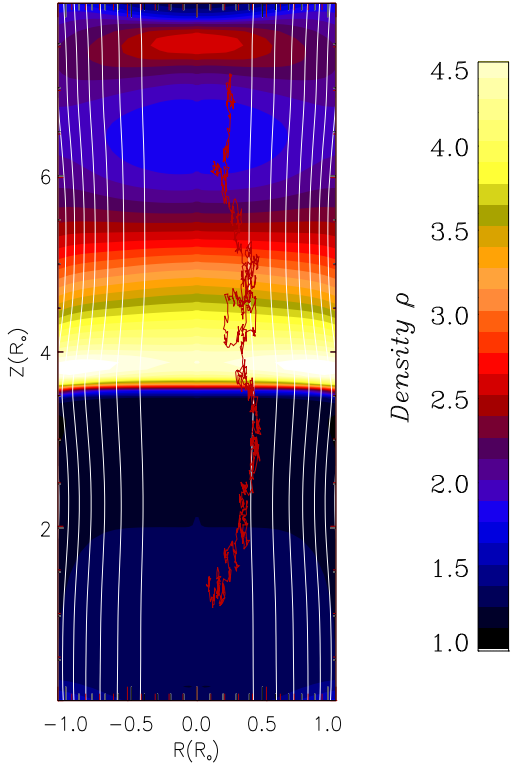}}
\caption{Zoom in a jet snapshot  where Kelvin-Helmholtz instabilities
are active. The parameters of the MHD simulations are the same as in
Fig.\ref{f2}. The grey levels represent the density levels while the white
lines are magnetic surfaces. A shock arises in the core of the jet ($R\leq
1$) with a plane shape perpendicular to the jet axis. Using a large number
of particles like the one which trajectory is displayed with a thick white
line, we measure, in the shock frame, the stationary energetic spectrum of
particles at the shock front.}
\label{Shpart}
\end{figure}

\begin{figure}[t]
\resizebox{\hsize}{!}{\includegraphics{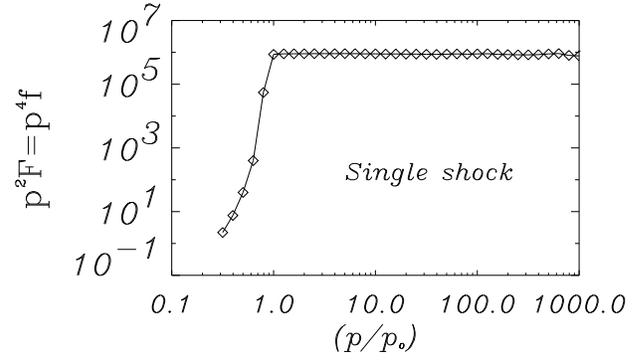}}
\caption{Energetic spectrum of particle population injected at momentum
$p_o$ in the MHD jet of Fig.~\ref{Shpart}. The spectrum is measured at the
shock front and is in a very good agreement with DSA theory predicting a
power-law of index $-4$ for a single plane shock of compression ratio
$r=4$. Note that in this computation the diffusion coefficients have
constant values fulfilling relation (\ref{Explicit2d}).}
\label{1Choc}
\end{figure}

\subsection{Realistic plane shock}
In this subsection we address the issue of the production of energetic
spectra by plane shocks arising from MHD simulations. This issue is a
crucial test for the relevance of SDEs using the velocity divergence 
defined in Eq.(\ref{div}). 
We stress that {\it all simulations performed in this paper are done 
using test-particle approximation}, i.e. no retroactive effects of the
accelerated particles on the flow are taken into account.

\subsubsection{Strong shock energetic spectrum}
\label{ssSsh}
We have performed a series of MHD simulations of cylindrical jets subject
to Kelvin-Helmholtz instabilities (cf. Sect.~\ref{mhdsimu}).  We selected
the case of a plane shock (quite common in the KH instability  simulations)
propagating along the jet with a radial extension up to the jet radius (see
Fig.~\ref{Shpart}). Its compression ratio is $r=4$ (measured by density
contrast) and constant along the shock front. We have chosen a particular
snapshot of the structure displayed on Fig.~\ref{Shpart}. By rescaling the
vertical velocity in order to be in the shock frame (where the down and
up-stream velocities are linked by $u_{\rm{down}}=u_{\rm{up}}/r$),  we
first consider this shock with infinite vertical boundaries and reflective
radial boundaries. Namely, we set that  if the particle is escaping the
domain at $Z<Z_{min}=0$ or $Z>Z_{max}=8$, we take the velocity to be
$\UU_{p}(Z>Z_{max})=\UU_p(R,Z_{max})$ (same thing for $Z< Z_{min}$). The
condition allows for particles far from the shock to eventually return and
participate to the shaping of $F(p)$.  The reflective radial boundaries are
located at the jet axis $R=0$ (to avoid the particle to reach $R=0$ where
SDEs are not valid) and $R=1$. Such boundaries ensure that no particle can
radially escape from the jet during the computation. The constant value of
the diffusion  coefficients $D_{ZZ}$ and $D_{RR}$ must fulfill relations
(\ref{Explicit2d})  and (\ref{Minimum}). Actually, in the particular case
of a plane shock propagating along the vertical axis, only $D_{ZZ}$ must
fulfill previous relations, namely $D_{ZZ} > D_{min}=X_{sh}|u_Z|/2$. The
shocks width $X_{sh}$ is defined as the location of the most negative
velocity divergence of the flow. Typically, this width corresponds to the
size of a mesh cell in the case of strong shock. We can then safely set
$D_{ZZ}=0.4$ as we will have $D_{ZZ} = 10 D_{min}$. The radial diffusion
coefficient is tuned as $D_{RR} =0.01$ and will enable particle to explore
the shock front structure. On Fig.~\ref{1Choc} we display the results of
the use of SDEs on a particle population injected at momentum $p=p_o$ and
propagating in snapshot represented by Fig.~\ref{Shpart}. We easily see
that the resulting spectrum is a power-law of index $-4$ completely in
agreement with DSA theory (see Section \ref{Sacc}). The existence of a few
particle 
with $p<p_o$ arises from the fact that outside the shock, the velocity
divergence is not equal to zero, as it would be with a prescribed velocity
profile \citep{Krul94,Marc99}. Note that in the absence of other energetic
mechanism (as second-order Fermi acceleration or synchrotron losses), the
simulation is independent of the physical value of $p_o$ as the diffusion
coefficient is independent of $p$.

\subsubsection{Single shock with synchrotron losses} 

\begin{figure}[t]
\resizebox{\hsize}{!}{\includegraphics{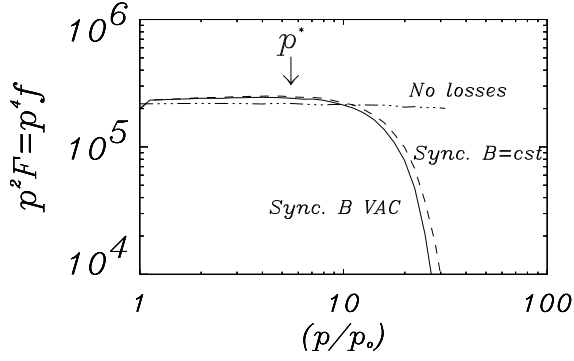}}
\caption{Energetic spectrum of energetic population injected at momentum
$p_oc=100MeV$ in the MHD jet of Fig.~\ref{Shpart} and subject to synchrotron
emission. The spectrum is measured at the shock front. It matches the
solution of 
\citet{webb84}, in particular for the cut-off momentum $p^*$ where synchrotron
losses balance shock energy gains. For this computation, we have considered
magnetic field given by the MHD code. The upstream
velocity $U_{up} = 300 \rm{km/s}$, and the mean magnetic field $B_o= 100 \ \rm{\mu G}$.}
\label{Sync}
\end{figure}
For electrons, the
acceleration occurring within shock may be balanced at the cut-off by
radiative losses due to the presence of the jet magnetic
field. \citet{webb84} has presented a complete analytical resolution of
Fokker-Planck transport equation including both first-order Fermi
acceleration and synchrotron emission. In particular, they show that the
energetic spectrum exhibits a cut-off at a momentum $p^*$ depending on
spatial diffusion coefficient and velocity of the flow. The choice of the
injection energy $p_oc$ of electrons is determined by the lower boundary of
the inertial range of magnetic turbulence. Indeed, to interact with
turbulence and to spatially diffuse, electrons must have momentum larger
than $p_i= m_iV_A$, where $m_i$ is typically the proton mass and $V_A$ is
the Alfv\'en speed \citep{lacom77}.  The energy threshold corresponds to
\beq \epsilon_i=p_ic \simeq 900 MeV \left(\frac{V_A}{c}\right)
\label{synch1}
\eeq
\noindent In our simulations, we assume an Alfv\`en speed $V_a \sim 2.2 \
10^8 \ \rm{cm/s} \sim c/100$ (see the discussion in Section \ref{SSDA})
leading to $p_oc= 100 MeV\ge p_ic$. As previously noted, the Alfv\`en
speed in extragalactic jets can reach appreciable fraction of the light
speed. An increase  of $V_a$ leads to an increase of the particle injection
threshold and a decrease of the dynamical  momentum range explored. In that
case, the Fermi second order effect must be included in our SDE system (via
the diffusive  term in momentum in Eq.~\ref{SDE7}). Time dependent
simulations (in progress) will include the associated discussion.\\ 
The
result of the simulation including synchrotron losses is displayed on
Fig.~\ref{Sync}.  When assuming a constant magnetic field and diffusion
coefficients,  the cut-off energy $\epsilon^*=p^*c$ reads as \citep{webb84}
\begin{eqnarray} 
\epsilon^*&=&\frac{m_e^2c^3}{D_{ZZ}}\frac{2\pi}{\sigma_TB^2}u_{\rm{up}}^2\frac{r-1}{r(r+1)}\nonumber\\
&=&0.48 GeV\left(\frac{U_{\rm{up}}}{300 km/s}\right)\left(\frac{R_{jet}}{100
pc}\right)^{-1}\left(\frac{B}{100\mu G}\right)^{-2}
\label{sunc2}
\end{eqnarray}      
\noindent Fig.~\ref{Sync} displays the spectrum at the shock in case of a
the magnetic field  obtained from the MHD code. The cut-off is in good
agreement  with the resulting 
spectrum despite the numerical simulation is considering a spatially
varying magnetic field. The Fig. ~\ref{Sync} also shows  the case of a
constant magnetic field taken as the average of the previous one.

\begin{figure}[t]
\resizebox{\hsize}{!}{\includegraphics{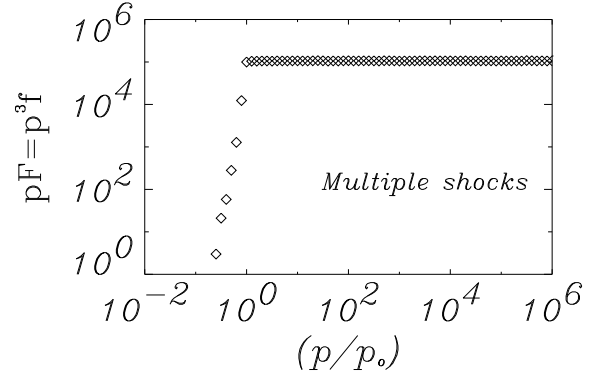}}
\caption{Energetic spectrum resulting from acceleration by multiple shocks
in an extragalactic jet. The power-law arising from this simulation matches
exactly the 
result given by DSA theory in the case of no particle escape from the shock
region ($t_{esc}\gg t_{acc}$). To achieve this simulation, we have
considered the snapshot in Fig.\ref{Shpart} but with vertical
re-injection of escaping particles. These boundaries mimic the effect of
multiple shocks interaction with particle during their propagation along
the jet.}
\label{MChoc}
\end{figure}

\subsubsection{Multiple shocks acceleration}
\label{Smsa}
\begin{figure}[t]
\resizebox{\hsize}{!}{\includegraphics{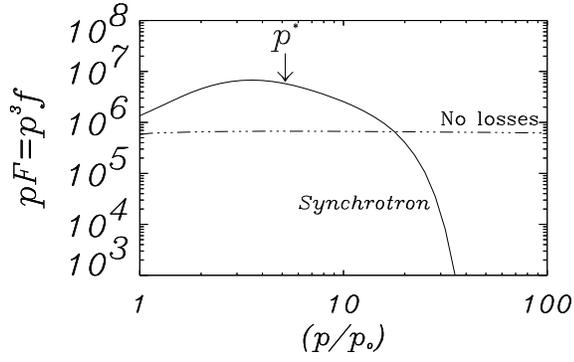}}
\caption{ Momentum spectrum of electrons injected at $p_oc=100 MeV$ in a
jet prone to multiple shocks acceleration. Note the bump occurring because
of the synchrotron cut-off that tends to accumulate particles at the
cut-off momentum $p^*$. }
\label{MZchoc}
\end{figure}

The presence of multiple shocks increases the efficiency of particle
acceleration. In multiple shocks, the particles accelerated at one shock are advected 
downstream towards the next shock. The interaction area is enhanced, so as the
escaping time. The general expression of the distribution function at
shocks front, $\log f\propto -(3+t_{acc}/t_{esc})\log p$ will then tend to
$\log f \propto -3\log p$. This multiple shocks acceleration may occur in
jets where numerous internal shocks are present \citep{FM97}. We intend to
modelize this effect using the same snapshot as in previous calculations
but changing the nature of the vertical boundaries. Indeed, since we are
modelizing only a small part of the jet (typical length of $800 pc$), we
can assume that if a particle is escaping by one of the vertical
boundaries, it can be re-injected at the opposite boundary with identical
energy. The re-injection mimics particle encounters with several parts of
the jet where shocks are occurring. Physical quantities are set to same
values than in paragraph dealing with single plane shock. The result of the
simulation is displayed on Fig.~\ref{MChoc} and on Fig.~\ref{MZchoc} when
synchrotron losses are considered. On Fig.~\ref{MChoc}, the spectrum
reaches again a power-law shape but with a larger index of $-3$, consistent
with previous statements. When synchrotron losses are included in SDEs
(Fig.~\ref{MZchoc}), we find a similar spectrum than for single shock but
with some differences. Namely, the curve exhibits a bump before the
cut-off. This bump can easily be understood since the hardening of the
spectrum enables particles to reach higher energies where synchrotron
losses become dominant. Thus an accumulation of particles near the cut-off
momentum $p^*$ will occur. The bump  energy corresponds to the equality
between radiative loss timescale and multiple shock acceleration
timescale. The last timescale is larger than the timescale required to
accelerate a particle at one isolated shock because of the advection of
particles from shock to next one \citep{Marc99}.

\begin{figure*}[t]
\centering
\includegraphics[width=17cm]{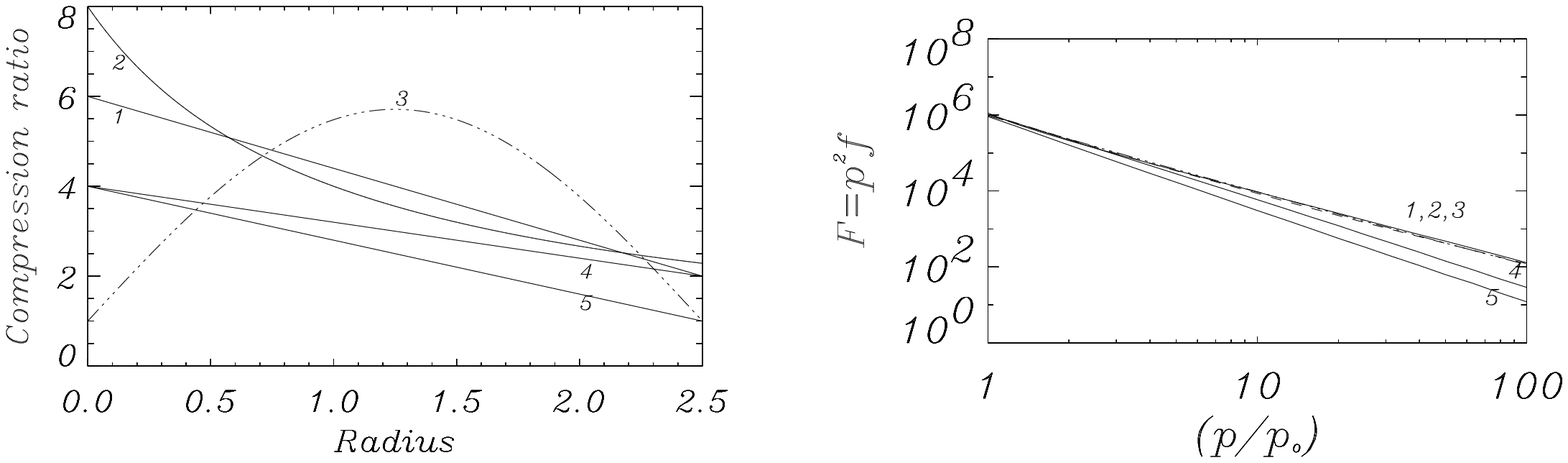}
\caption{Spectrum produced by plane shocks with spatially varying
compression ratio $r$. The left panel represents compression ratio profiles
along the shock front while the right panel displays the resulting
spectra. The calculations 1, 2 and 3 have different $r-$profiles but the
same $r_m=4$. The resulting spectra are corresponding to curve 1,2 and 3 on
the right figure. We can see that this curves are almost the same and are
very close to a power-law with index $-(r_m+2)/(r_m-1)$. Curves 4 and 5
correspond to compression ratio profile with mean values $r_m$ equal to
$3.5$ and $3$. The corresponding spectra on the right figure are again
consistent with power-law with indices controlled by their $r_m$. All particles are
injected at $p_o c = 100 \ \rm{MeV}$ along the shock.}
\label{Thvar}
\end{figure*}

\section{Acceleration at complex shock fronts}
\label{SNI}
The shock structures are subject to important evolution during the development 
of the KH instability. We now investigate the particle distribution function
produced at these shocks using the SDE formalism. All the shock acceleration process
here is investigated using snapshots of the MHD flow.

\subsection{Plane shocks with varying compression ratio} 
\label{Sdist}
In astrophysical and particularly jet environments, (weak) shocks occurring
within magnetized flows in the {\it early phase} of KH instability (see
Fig.~\ref{f2}) are non-planar and/or with  non-constant compression ratio
along the shock surface. We first consider analytical calculations of the particle
distribution produced in such shocks that extend previous works and we complete our estimates 
using the MHD-SDE system. 

\subsubsection{Analytical approach}
The theory of DSA explains the energetic spectrum of diffusive particles
crossing plane shock with constant compression ratio $r$. Even when the
plane structure is relaxed (\cite{D83}) the compression ratio is usually
assumed as constant along the shock front. In astrophysical jets, complex
flows arise from the jet physics so that even the plane shock assumption
is no longer valid implying a non-analytical derivation of the particle distribution
function. Nevertheless, it seems obvious that if the shock front is not
strongly bent, the particle acceleration process should not be strongly modified.\\
Let us first quantify this assertion. We calculate the mean momentum 
gained by a particle during one cycle (downstream $\rightarrow$ upstream $\rightarrow$
downstream) 
\beq \frac{<\Delta p>}{p}=
\frac{4}{3}(r(R)-1)\frac{u_d}{v}
\label{NPS1}
\eeq
\noindent we assume that, during this cycle, the particle sees the
local structure of the shock as a plane ($v$ is the speed of the
particle), e.g. the spatial scale where the shock bends is large 
compared to the particle diffusive length. \\
\noindent Here, contrary to the standard DSA theory, the energy gain
depends on the location of the shock crossing of the particle. The probability for a
particle to escape from the shock during one acceleration cycle is however still
given by the usual DSA theory, namely $\eta_k=4u_d/v_k$ ($v_k$ is the
speed of the particle during the $k$th cycle). The probability
that a particle stays within the shock region after $n$ cycle $Pr_n$ can be
linked to the mean  momentum gain after $n$ cycle as
\beq
\frac{\ln{Pr_n}}{\ln{p_n/p_o}}=\frac{\sum_{k=1}^{n}\ln
(1-\eta_k)}{\sum_{k=1}^n\ln (1+\eta_k(r_k-1)/3)}
\ .
\label{NPS2}
\eeq
\noindent The compression ratio $r$ depends on the number of
the cycle since in reality, the particle is exploring the front shock
because of the diffusion occuring along the shock front. This expression
can be simplified if we assume the flow background velocity very small
compared to particle velocity ($\eta_k\ll 1$ for a non-relativistic
shock) and that particles are ultra-relativistic ($\eta_k=\eta=4u_d/c$). 
The expression then becomes 
\beq
\frac{\ln{Pr_n}}{\ln{p_n/p_o}}\simeq -3\frac{n}{\sum_{k=1}^nr_k - n} \ .
\label{NPS4}
\eeq
\noindent The sum of the different compression ratios experienced by
particle population can be approximated using the average compression ratio
measured along the shock front $r_m$. Indeed, each particle interacting with
the shock are prone to numerous cycles of acceleration and then the sum
remaining in Eq.(\ref{NPS4}) can be expressed as $\sum_{k=1}^nr_k\simeq
nr_m$. Hence, the energetic spectrum is a power-law but with an
index controlled by the mean value of the compression ratio all over the
shock front, namely 
\begin{figure}[t]
\resizebox{\hsize}{!}{\includegraphics{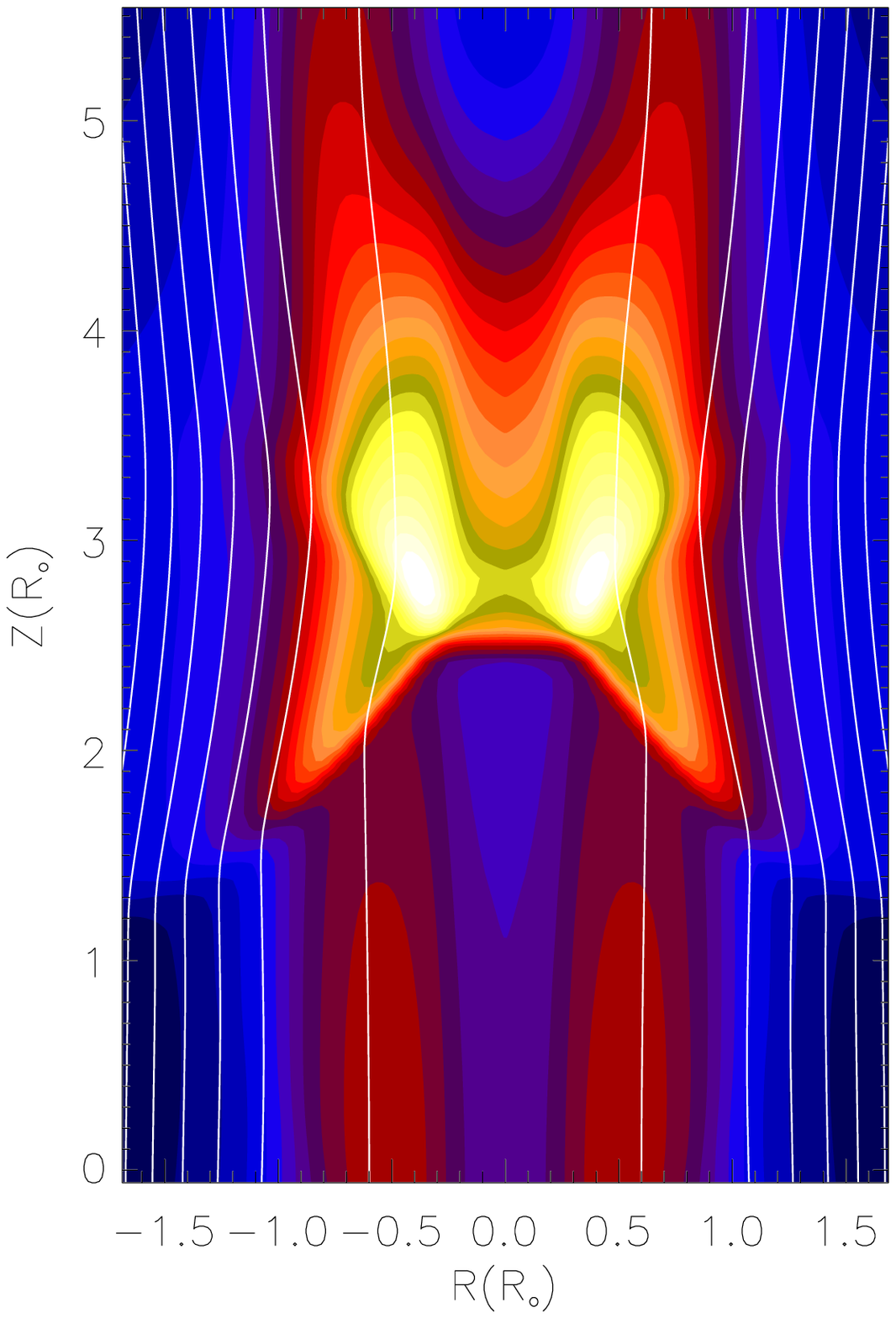}}
\caption{Same plot as Fig.(\ref{Shpart}). The MHD simulation parameters
are the same as in Fig.\ref{Shpart} except for $\delta M_o=0.5$. This
snapshot is selected during the developing phase of the shock where the
structure is evolving toward a plane shock. The inclined (with respect to
the jet axis) part of the shock front are affected by a strong magnetic
bending due to magnetic conservation through the shock.}  
\label{Oblik}
\end{figure}
\beq
p^2f\propto \frac{\p\ln Pr_n}{\p p_n} \propto p^{-3(r_m+2)/(r_m-1)} \ .
\label{NPS6}
\eeq
\noindent In this demonstration, the compression ratio profile
itself is not involved in the spectrum shape but only its average
value $r_m$, as long as one can consider the shock to be locally plane.    
Eq.(\ref{NPS6}) generalizes the results provided by \citet{D83} concerning 
curved shocks with constant compression ratio. If the plane shock assumption is relaxed, 
numerical simulations are necessary.\\
In order to complete this result, we have performed several numerical
calculations where a mono-energetic population of relativistic particles are
injected with momentum $p_o$ behind an analytical prescription describing a
plane shock with varying compression ratio (the shocks are test examples). 
The result of this numerical test is displayed on Fig.(\ref{Thvar}).\\
\noindent In this test, we have done three calculations with three
different  compression ratio profiles (curves 1, 2 and 3) but with
identical average  value $r_m=4$. Setting both vertical and radial
diffusion, we have obtained the spectra 1, 2 and 3  displayed on right
panel of Fig.(\ref{Thvar}). These three curves are almost the same.  On two
other calculations, we have chosen linear profiles with different  values
of $r_m$ (curves 4 and 5): again a power-law spectrum is found with indices
consistent  with previous analytical statements. This conclusion is correct
only if during on cycle the particle mean free path along the shock front
is small compared to its curvature and if during many cycles the particle
is able to explore the whole shock structure.

\subsubsection{Locally-plane shock}
\begin{figure*}[t]
\centering
\includegraphics[width=17cm]{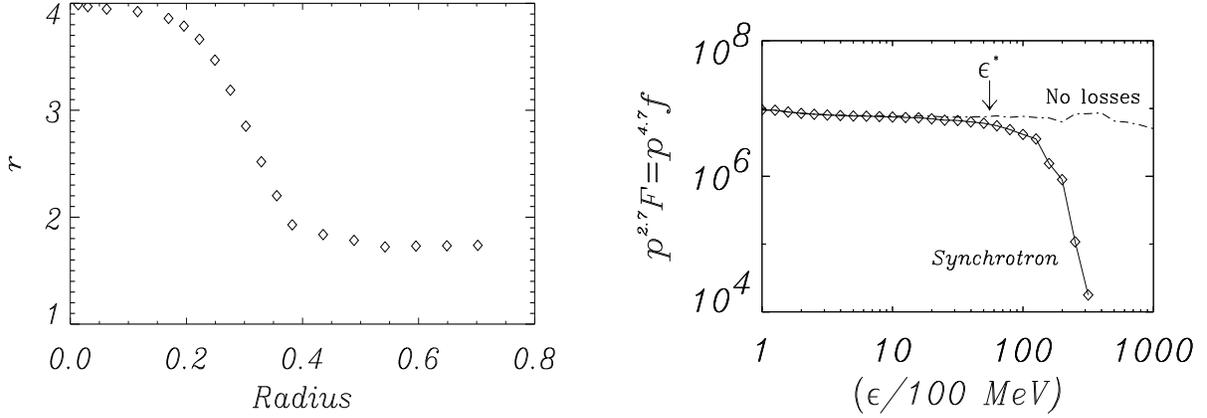}
\caption{Compression ratio profile (left) and energetic spectrum (right) of
accelerated particles by the shock displayed on Fig.\ref{Oblik}. This shock
is not a plane shock and does not have constant compression ratio along its
shock front. The curvature radius of this shock is much larger than the
mean free path of a particle so, locally, the shock can be considered as a
plane shock. The resulting spectra (with only acceleration or with
synchrotron losses included) are consistent with a plane shock with
compression ratio $2.7$ which is close to the average value of the
compression ratio of this shock, namely $r_m=2.68$.}
\label{Revar}
\end{figure*}

The previous considerations can be applied to a non-planar shock produced
in the early  stage of the axisymmetric Kelvin-Helmholtz instability. The
inner shocks tend to evolve from  curved fronts in the early phases of the
instability toward plane shocks, perpendicular to the jet axis  (see
Sect.\ref{Injet} and Fig.\ref{f2}). On Fig.\ref{Oblik} the curvature radius
of the shock  is typically of the order of the jet radius while the
obliquity angle (between the shock front  and the jet axis) ranges from
zero to $27^o$. For such a low obliquity the shocks are subluminal.  A more
subtle consequence of the non-constancy of the compression ratio is that
the electric fields  generated along the shock front cannot be canceled by
any Doppler boost. In other words,  complex shocks do not have a unique de
Hoffman-Teller frame. This problem strongly complexifies the particle
acceleration and transport in jets and is postponed to future works
especially treating  strongly oblique (or even perpendicular) shocks. In
the present paper, the MHD shocks are only  weakly oblique and
non-relativistic (the effects of electric fields on particle acceleration
are neglected).  In principle, once the electro-magnetic field is known
throughout the jet, the systematic electric effects  on particle
trajectories can be implemented in the SDE system.\\
\noindent Keeping the same prescription for diffusion coefficients than in previous section (constant 
diffusion coefficients and radial reflective boundaries) we first have to verify the quasi-planar 
condition of the shock. To this aim, we form the ratio of the typical diffusion length occurring 
during one cycle along the shock front ($L_{\perp}$) and the curvature radius. 
The duration of one acceleration cycle is controlled by the residence time at the shock
$t_{res}=2D_{ZZ}/u^2_d$ (assuming that it is composed of $n$ identical
cycle). The number of cycle needed for the particle to escape the shock is
obtained when the escaping probability after $n$ cycles is equal to unity,
namely $\sum_{k=1}^n\eta_k=4nu_d/c$ when particles are relativistic. The
duration of one cycle is thus $\tau=8D_{ZZ}/u_dc$. The criterion for 
considering a shock as locally plane will be
\beq
\frac{L_{\perp}}{L_{curv}} \leq \frac{\sqrt{2D_{\perp,S}\tau}}{R_{jet}}
\label{NPS7}
\eeq
\noindent where $D_{\perp,S}$ is the maximal value of the diffusion
coefficient in the direction parallel to the shock front. With the
previously prescribed diffusion coefficients the ratio has a maximum value
equal to $3.4\ 10^{-2}$. This value is small compared  to unity which means
that during one cycle, the particle will interact with a zone of the shock
where the compression ratio is almost constant. On the other hand, this
ratio is not so small and within a few cycles of acceleration particles
will explore a significant part of the shock front.\\
\noindent Figure(\ref{Revar}) shows the energetic spectrum produced in such
curved shock. The result is close to a power-law of index $f\propto
p^{-4.7}$  and when synchrotron losses are taken into account, the cut-off
energy corresponds to the case of a uniform shock with constant compression
ratio equal to $2.76$. The cut-off,  given by Eq.(\ref{sunc2}) is close to
the value obtained on the plot. We postpone to Section \ref{onechoc} the
comparison between particle acceleration timescale and shock survival
timescale in the different phases of the jet evolution. We can however
anticipate here that for typical jet parameters the former is smaller than
the latter. This validates our results obtained using MHD snapshots.\\

\subsection{Strong shock acceleration and spatial transport}
\label{Splane}
We now consider the shock acceleration and spatial transport in chaotic magnetic field in 
strong shocks occurring in the {\it late phase} of the KH instability where the most efficient 
particle acceleration is expected \citep{Mietal99}. The validity of the snapshot approach
is tested against the survival of the shocks. We investigate the effect of 
radial escape on the particle distribution in the single and the multiple shock configuration.

\subsubsection{Maximum energies expected and electron transport}
\label{Sana}
\noindent The maximum electron energy is limited by radiative or escape
losses. In case of synchrotron radiation, the loss timescale is
$\tau_{\rm{loss}} \sim 1.2 \ 10^4 (B_{\rm{mG}})^{-2} \ E^{-1}_{GeV} \ \rm{yr}$
which compared to the acceleration timescales presented above leads to
electron  with energies $\gamma_{max} \sim 10^8 \ (U_{jet}/c)^{3/2}$, around 1
TeV for $U_{jet}/c=0.1$ (the magnetic field and the particle energy are
expressed in mGauss and in GeV units respectively).\\ 
In a quite general way, the radial and vertical diffusion coefficients can be written as
\begin{eqnarray}
D_{RR}&=&D_{\parallel}\left|\frac{B_R}{B}\right|+D_{\perp}\left(1-\frac{B_R^2}{B^2}\right)^{1/2}\nonumber\\
D_{ZZ}&=&D_{\parallel}\left|\frac{B_Z}{B}\right|+D_{\perp}\left(1-\frac{B_Z^2}{B^2}\right)^{1/2} 
\label{Radloss1}
\end{eqnarray}
\noindent where $B$ is the total magnetic field amplitude. The confinement
time is driven by the radial diffusion coefficient $D_{RR}$ which may be
expressed in term of $\eta_T$ as
\beq
D_{RR}\simeq\frac{D_o}{\eta_T}(\alpha + (1-\alpha^2)^{1/2}\eta_T^{2.3})\ ,
\label{Radloss2}
\eeq
\noindent where $D_{\parallel}=D_o/\eta_T$ and $\alpha$ stands for the
average value of $|B_R/B|$ all over the simulation box. The coefficient
$D_o$ may eventually depends on the particle momentum. It can easily be
seen that Eq.(\ref{Radloss2}) has a minimum value for  
\beq
\eta_T^{min}=\left(\frac{\alpha}{1.3 \ (1-\alpha^2)^{1/2}}\right)^{1/2.3} \
.     
\label{Radloss3}
\eeq
\noindent At $\eta_{min}$, particle confinement reaches its
maximum (see Eq.~\ref{test3}). Typically, radio jets do not display opening
angle larger than a 
few degrees leading to $\eta_T^{min}$ of the order of $0.2$. Paradoxically,
low turbulence levels do not provide efficient confinement since
largest diffusion motion occurs along the magnetic field which have locally
radial components. Within a timescale $\tau_{loss}$ electrons are able to
explore distances   
\beq 
\Delta R = \sqrt{4 \ D_{RR} \ \tau_{loss}} \ .  
\eeq 
\noindent In the chaotic magnetic regime, Eq.(\ref{CDi1}) leads to 
\beq 
\Delta R \sim 15 \ [\frac{(\alpha + (1-\alpha^2)^{1/2}\eta_T^{2.3})}{\eta_T}]^{1/2} \ E_{\rm{GeV}}^{-1/3} \
B_{\rm{mG}}^{-7/6} \ \rm{pc}.
\label{drdiff} 
\eeq 
\noindent We consider a mean turbulence level $\eta_T = \eta_T^{min}$, and
assumed a magnetic field  $B = 100 \ \rm{\mu G}$, and a maximum turbulence
scale $\lambda_{max} \sim R_{\rm{jet}}$.  For $\alpha \sim 2^o$, we get
$\Delta R(\rm{1 \ TeV}) \sim 9 \ \rm{pc}$ and $\Delta R(\rm{1 \ GeV})
\sim 90 \ \rm{pc}$. The high energy  electrons are only able to explore about
one tenth of the jet radius and can be considered as confined to the
region where they have been injected. The GeV electrons can explore larger
fraction of the jet and escapes are expected to steepen the particle
distribution. These are averaged results, $\Delta R$ is sensitive to the
magnetic field, for example if  B decreases (increases) by one order of
magnitude $\Delta R$ increases (decreases) by a factor $\sim 14$. Along
the jet, particles are advected from one shock to the next on timescales
$\Delta Z_{shock}/V_{flow}$, where a mean inter-shock  distance $\Delta
Z_{shock} \sim 1 \ \rm{kpc}$ and $V_{flow} \sim 10^{-2 / -1} c$ lead to
$\tau_{adv \parallel} \sim 3 \ 10^{4 / 5} \rm{yr}$.  The high energy part 
of the electron distribution is then produced by one shock and can hardly 
be re-accelerated in a second one downstream. The spectrum at these energies 
strongly depends on the shock compression ratio. At lower energies GeV electrons 
distribution can be subject to either transversal escapes or multiple shock effects.
 For both populations, the electrons accelerated at inner shocks remain within
1 kpc from their injection points, this clearly separates the inner jet to the Mach
disc and justifies a fortiori our approach simulating only the kiloparsec scale jet.
It also clearly appears that the spatial transport issue addresses to the multi-wavelength
morphologies of jets. We know make these statements more precise using the coupled MHD-SDE system.

\subsubsection{Single shock}
\label{onechoc}
\begin{figure}[t]
\resizebox{\hsize}{!}{\includegraphics{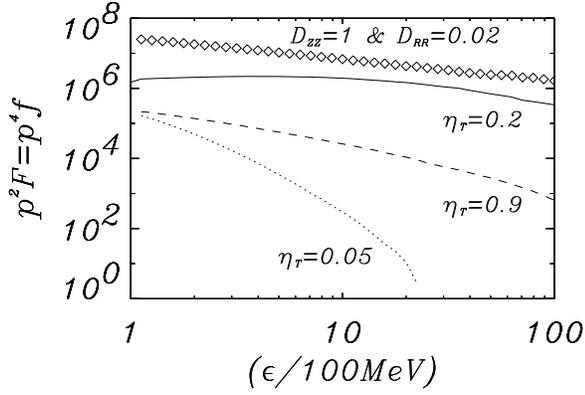}}
\caption{Spectra produced at the shock front displayed on Fig.\ref{Shpart}
without outer reflective boundary, namely with radial particle losses. The
upper plot represents a spectrum done with constant diffusion coefficients
($D_{ZZ}=1$ and $D_{RR}=0.02$). The radial losses modify the spectrum by
increasing the index of the power-law from $-4$ to $-4.25$. The last three
curves are spectra produced by using realistic diffusion coefficients given
by Eq.(\ref{CDi1}). The momentum dependence of these coefficients modifies
the shapes that are no longer power-laws (see Sect.\ref{onechoc}). The
upper plot has an arbitrary normalization unrelated to the three last
curves.}   
\label{Dreal}
\end{figure}

 So far, we have presented numerical calculations using reflective radial
boundaries (no particle losses) and constant diffusion coefficients. 
In this section, we choose to remove step by step these two constraints.
Starting from the snapshot of Fig.\ref{Shpart}, we first remove
the outer reflective boundary and consider any particle
having $R>R_{jet}$ as lost. Then we adopt diffusion coefficients given by
Eq. (\ref{CDi1}) since they  
arise from a transport theory consistent with high turbulence levels
$\eta_T$ and are confirmed numerically. Quasi-linear theory does likely
apply at very low turbulence levels implying high parallel diffusion
coefficient and acceleration timescales. Expected spectra must then be softer
than the same spectrum obtained in chaotic regime. \\
\noindent First, as an illustration of escape effects, we consider the case
of constant diffusion  coefficients, namely $D_{ZZ}=1$  and
$D_{RR}=2.10^{-2}$. The resulting spectrum can be seen on Fig.\ref{Dreal}
and is consistent with a harder power-law. In previous simulations, the
escaping time was defined as the time needed by the flow to advect a
majority of RPs away from the shock. Here the effect of the confinement
inside the jet if lower than the escaping time from the shock will be the
main source of particle losses. The distribution function reads as
$\log{f}\propto -(3+t_{acc}/t_{loss})\log{p}$ where $t_{loss}=
min(t_{conf},t_{esc})$ and will stay as a power-law as long as the escaping
time is not momentum dependent. In our example
$t_{loss}=t_{conf}=(R_{jet}-R_{inj})^2/4D_{RR}=10.125$, where $R_{inj}$ is
the average radius of injected particles. The resulting  spectrum index on
Fig.\ref{Dreal} is in good agreement with this  estimate since the ratio
$t_{acc}/t_{conf}=6D/(t_{conf}u_d^2(r-1))=1.26$ and the plot representing
the spectrum done with these constant  diffusion coefficients has a
power-law index as $f\propto p^{-4.25}$.\\  Secondly, we discuss the case
of Kolmogorov turbulence and keep $\eta_T$ free in order  to check its
influence on the transport of particles. The three last plots in
Fig.\ref{Dreal} represent simulations performed without outer reflective
boundaries and diffusion coefficients as described by
Eq.\ref{Radloss1}. The  simulations account for energy as well as spatially
($B_r$ and $B_z$ are both  function of r and z) dependent transports. Each
curve corresponds to a value of  the turbulence level $\eta_T=0.05, 0.2$
and $0.9$.  In a Kolmogorov turbulence  $D_{\parallel}\propto D_{\perp}
\propto p^{1/3}$, $t_{conf}\propto  D_{RR}^{-1}\propto p^{-1/3}$ which
leads to a confinement time decreasing   while increasing momentum and a
convex spectrum. At a low turbulence level,  the ratio $t_{acc}/t_{conf}$
is large, increases with the particle momentum and  leads to softer spectra
with low energy cut-off (at few GeV/c). In order to get  significant
particle acceleration and large energy cut-off (beyond 1 TeV) turbulence
levels  $\eta_T \ge 0.1$ seem mandatory. The maximum confinement is
obtained for turbulence level  $\eta_T$ compatible with Eq.
(\ref{Radloss3}).\\
\noindent One important issue to discuss about is the validity of our
results while considering snapshots produced from the MHD code. It appears
from Fig.~\ref{f2} that both weak curved and strong plane shocks survive a
timescale of the order of $ 5 \ \tau_o$. The shock  acceleration timescale
of a particle of  energy $E_{\rm{GeV}}$ may be expressed as $\tau_{acc}
\sim 20 \ D_{ZZ}/U^2_{\rm{sh}}$ for  a compression ratio of 4 \citep{D83},
where $U_{\rm{sh}}  \sim 10 \ c_s$  is the shock velocity. Using the
Eq.~\ref{CDi1}) and \ref{Radloss1} we end up with a typical ratio
$\tau_{acc}/\tau_0 \sim 10^{-2} \ \eta_T^{-1} \ E^{1/3}_{\rm{GeV}} \
B^{-1/3}_{mG}$. Our snapshot then describes well the shock acceleration
(e.g. $\tau_{acc}/\tau_0 \le 1$) up to TeV energies unless the turbulence
level is very low and the magnetic field much lower than $100 \mu G$. The
conclusion is the same for curved shocks as the acceleration timescale is
smaller in that case.  However, time dependent simulations are required to
a more exhaustive exploration of the jet parameter space  and to test the
different turbulence regimes.

\subsubsection{Multiple shock-in-jet effects}
\begin{figure}[t]
\resizebox{\hsize}{!}{\includegraphics{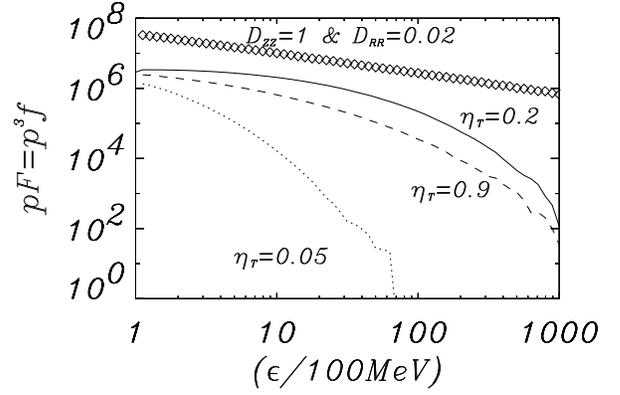}}
\caption{Same plot as in Fig.\ref{Dreal} but with periodic vertical
boundaries. This setting mimics the effect of multiple shocks
acceleration. As previously, inclusion of radial particle losses
affects spectra: a softening of the spectral index, cut-off energies 
dependent on $\eta_T$.}
\label{RealM}
\end{figure}
\begin{figure}[t]
\resizebox{\hsize}{!}{\includegraphics{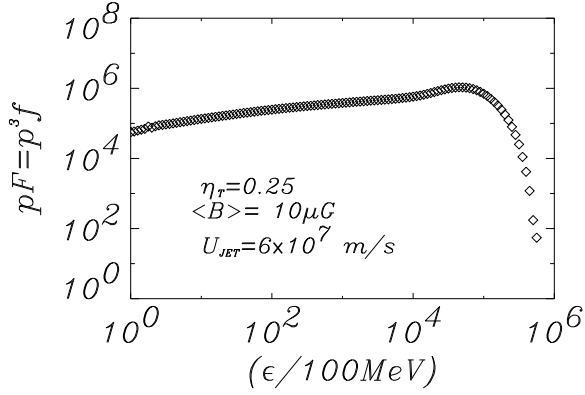}}
\caption{Multiple inner-jet shocks spectrum including synchrotron losses for
$\eta_T=0.2$, $B_o=10\mu G$ and $U_{jet}=c/5$. The synchrotron cut-off
energy lies at a few $TeV$. The most energetic electrons at the bump of this
distribution would have a peaked synchrotron emission at $\nu\sim 1.9\
10^{16}Hz$ which corresponds to UV/X-ray emission ($h\nu\sim 20 eV$).}
\label{RSyn}
\end{figure}

The radial losses should also affect the transport of particles
encountering several shocks during their propagation. This description is
pending to the possibility of multiple strong shocks to survive few
dynamical times. This issue again requires the time coupling between SDE and MHD simulations
to be treated.\\
However, the general statement about the distribution function is still valid but at the
opposite of previous multiple shocks acceleration calculations (see Section
\ref{Smsa}) the lack of confinement is the only loss term. On
Fig.\ref{RealM}, we have performed the same calculations as in the
previous paragraph except that we impose periodic vertical boundaries where
particles escaping the computational domain by one of the vertical
boundary is re-injected it at the opposite side keeping the same energy. \\
We again start with our fiducial case displaying the spectrum obtained from
calculations done with constant diffusion coefficients, i.e. $D_{ZZ}=1$ and
$D_{RR}=2 \ 10^{-2}$ (the upper plot). The power-law index is modified  and
equals to $-3.13$ instead of $-3$ as obtained in calculations without
radial losses. This result is close to the analytic estimate since
$t_{acc}/t_{conf}\simeq 0.11$  in that case. In the chaotic diffusion
regime the same behavior is observed  in the spectra, e.g. convex shape,
low energy cut-off at low turbulence levels.  In this diffusion context,
multiple shock acceleration is again most efficient  for $\eta_T \sim
0.2-0.3$ and tends to produce hard spectra up to 10-100 GeV for  electrons
without radiative losses. The spectrum cut-off beyond 10 TeV.\\
In Fig.\ref{RSyn}, we have included synchrotron losses effects in one of the most
favorable case ($\eta_T=0.25$) in the chaotic regime. The resulting
spectrum shows a characteristic bump below the synchrotron cut-off lying
around a few ten GeV. This hard spectrum may be intermittent in jets
as already noticed by \citet{Mietal99}. The spectrum and bump may also 
not exist because of non-linear back-reaction of relativistic particles 
on the shock structure (this problem require the inclusion of heavier 
particles in the simulation). Beyond the electrons loss their energy before 
reaching a new shock as discussed in Section \ref{Sana}. The magnetic field 
used is $10\ \mu G$ and suggests that higher values are apparently not 
suitable to obtain TeV electrons. The synchrotron peaked emission of the 
most energetic electrons of this distributions gives an idea of the 
upper limit of radiative emission achievable by this inner-jet shock. 
In a $10\mu G$ magnetic field, these electrons radiate UV/X-ray photons as
\citep{Rybi79} 
\beq
h\nu_{syn} =  0.29\frac{3h\epsilon^2eB}{\mu_o m^3_ec^5}\sim 20 eV
\left(\frac{\epsilon}{10TeV}\right)^2\left(\frac{B}{10\mu G}\right)
\label{Syn1}
\eeq 
The electron population computed here does not go beyond $50 \ \rm{TeV}$,
which then suggests an energy emission upper limit around $\sim 0.5 \rm{keV}$.
The maximum energy scales as $U_{jet}$ (see Eq.(\ref{sunc2})) and can be
significantly increased in case of fast jets (with speeds up to c/2 the limit of
the validity of the diffusion approximation).\\
In case of inefficient confinement, e.g. $\eta_T$ different from 0.2-0.3,
this result also suggests that the synchrotron model may in 
principle not account for the X-ray emission of extragalactic jets 
probably dominated by another radiative mechanism (for instance the 
Inverse Compton effect). However again, we cannot draw any firm conclusions 
about this important issue and postpone it to the next work treating 
full time dependent simulations.
\noindent 

\section{Concluding remarks and outlook}
\label{Scon}
In the present work, we performed 2.5D MHD simulations of periodic parts of
extragalactic jets prone to KH instabilities coupled to a kinetic scheme 
including shock acceleration, adiabatic and synchrotron losses as well 
as appropriate spatial transport effects. The particle distribution 
function dynamics is described using stochastic differential equations 
that allow to account for various diffusion regimes.\\ 
We demonstrate the ability of the SDEs to treat multi-dimensional 
astrophysical problems. We pointed out the limits ($D_{min}$ defined 
in Eq.~\ref{Minimum})  imposed by the spatial resolution of the shock 
on the diffusion coefficient. The SDEs are applicable to
particular astrophysical problem provided $D \ge D_{min}$. We perform
different tests in 2D showing consistent results between numerical
simulations and  analytical solutions of the diffusion-convection equation.
Finally we demonstrate the ability of the MHD-SDE system to correctly
describe the shock acceleration process during the evolution of the KH
instability. Complex curved shock fronts with non constant diffusion
coefficients that occur at early stage of the instability 
behave like plane shock provided the diffusion length is smaller than the shock
curvature. The equivalent plane shock has a compression ratio equals to the
mean compression of the curved shock. In case of strong plane shocks which
develop at later stages of the KH instability, we found that the inclusion
of {\it realistic} turbulent effects, e.g. chaotic magnetic diffusion lead
to complex spectra. The resulting particle distributions are no more
power-laws but rather exhibit 
convex shapes linked to the nature of the turbulence. In this turbulent
regime, the most  efficient acceleration occurs at relatively high
turbulence levels of the order of $\sim 0.2-0.3$. The electron maximum
energies with synchrotron losses may go beyond 10 TeV for fiducial 
magnetic field values in radio jets of $\sim 10 \mu G$ and the spectrum
may be hard at GeV energies due to multiple shock effects.\\
However, in this work, SDEs were used on snapshots of MHD simulations 
neglecting dynamical coupling effects, preventing from any complete 
description of particle acceleration in radio jets. Such dynamical effects 
encompass temporal evolution of shock, magnetic field properties and 
particle distribution. The time dependent simulations will permit us to 
explore the parameter space of the turbulence and to critically test its 
different regimes.\\
The simulations have also been performed in test-particle approximation and 
do not account for the pressure in RPs that may modify the shock 
structure and the acceleration efficiency. This problem will be 
addressed in a particular investigation of shock-in-jet acceleration 
including heavier (protons and ions) particles.
Nevertheless the present work brings strong hints about the ability of 
first order Fermi process to provide energetic particles along the jet. 
Our first results tend to show that synchrotron losses may prevent 
any electron to be accelerated at high energies requiring either 
supplementary acceleration mechanisms or other radiative emission processes 
to explain X-ray emission as it has been recently claimed. Future works 
(in progress) will account for these different possibilities.

\acknowledgements
The authors are very grateful to E. van der Swaluw for careful reading of the manuscript
and fruitful comments, Rony Keppens and Guy Pelletier for
fruitful discussions and comments. A.M thanks J.G. Kirk for pointing him out the usefulness
of the SDEs in extragalactic jets. This work was done under Euratom-FOM Association Agreement with
financial support from NWO, Euratom, and the European Community's Human
Potential Programme under contract HPRN-CT-2000-00153, PLATON, also
acknowledged by F.C and partly under contract FMRX-CT98-0168, APP, acknowledged by A.M.
NCF is acknowledged for providing computing facilities.

\bibliographystyle{aa}

\end{document}